\documentclass[pre,superscriptaddress,showpacs,nofootinbib,floatfix,preprint]{revtex4-1}

\usepackage[utf8x]{inputenc}
\usepackage{graphicx}
\usepackage{amsmath}
\usepackage{amsfonts}
\usepackage{color}
\usepackage{epstopdf}
\usepackage{hyperref}

\newcommand{\cch}[1]{\left[#1\right]}

\newcommand{\prt}[1]{\left(#1\right)}
\newcommand{\aver}[1]{\left\langle#1 \right\rangle}

\hypersetup{colorlinks=true, citecolor=blue,
  pagecolor=blue,
  urlcolor=black,
  pdfcreator={pdflatex},
}

\begin{document}
\title{Lattice Model for water-solute mixtures.}

\author{A. P. Furlan\footnote[1]{email-furlan@if.ufrgs.br}}
\affiliation{Instituto de F\'{\i}sica, Unversidade Federal do Rio
  Grande do Sul, Caixa Postal 15051,91501-570,Porto Alegre, Rio
  Grande do Sul, Brasil}

\author{N. G. Almarza\footnote[2]{email-noe@iqfr.csic.es (deceased)}}
\affiliation{Instituto de Qu{\'\i}mica F{\'\i}sica Rocasolano,
  CSIC, Serrano 119, E-28006 Madrid, Spain }

\author{M.C. Barbosa\footnote[3]{email-marcia.barbosa@ufrgs.br}}
\affiliation{Instituto de F\'{\i}sica, Unversidade Federal do Rio
  Grande do Sul, Caixa Postal 15051, 91501-570, Porto Alegre, Rio
  Grande do Sul, Brasil}

\date{\today}
\begin{abstract}

A lattice model for the study of mixtures of associating liquids
is proposed. Solvent and solute are modeled by adapting the associating
lattice gas (ALG) model.
The nature of interaction solute/solvent is controlled by tuning
the energy interactions between the patches of ALG model. We have
studied three set of parameters, resulting on, hydrophilic, inert
and hydrophobic interactions.
 Extensive Monte Carlo simulations were carried out and the behavior
of pure components and the excess properties of the mixtures have
been studied.
The pure components: water (solvent) and solute, have  quite similar
phase diagrams, presenting:  gas, low density liquid, and high
density liquid phases. In the case of solute, the regions of
coexistence are substantially reduced when compared with both the
water and the standard ALG models.
A numerical procedure has been developed in order to attain series of
results at constant pressure from simulations of the lattice gas model
in the grand canonical ensemble.
The excess properties of the mixtures: volume and enthalpy as the
function
of the solute fraction have been studied for different interaction
parameters of the model. Our model is able to reproduce qualitatively
well the excess volume and enthalpy for different aqueous solutions.
For the hydrophilic case, we show that the model is able to reproduce
the excess volume and enthalpy of mixtures of small alcohols and
amines. The inert case reproduces the behavior of large alcohols such
as, propanol, butanol and pentanol. For last case (hydrophobic), the
excess properties reproduce the behavior of ionic liquids in aqueous
solution.
\end{abstract}
\maketitle
\section{Introduction}

Mixtures of water and organic solutes are of fundamental importance
for understanding biological and  chemical processes as well as
transport properties of fluids.
Even though the simplicity, of these solutions some of them show a
complex behavior of their thermodynamic and structural
properties~\cite{Evans1945}. For example, close to the  ambient
conditions, around $T=298.15 \:K$, $p = 1 {\rm bar}$, the excess
volume in binary mixtures of water and
alcohols~\cite{Marongiu1984,Pattel1985,Ben-Naim2008,Sha2016,Ott1993}
and of water and alkanolamines~\cite{Hepler1994,Stec2014} is negative
and it exhibits a minimum as the fraction of  the solute is increased.
 In the case of water-ionic liquids, however,  the excess volume
depend on the hydrophobicity of the solute. Simulation results suggest
 that for hydrophilic solutes as the  1,3-dimethylimidazolium chloride
the excess volume has a minimum as in the case of the alcohols,
whereas for  the case of more hydrophobic liquids as the
1,3-dimethylimidazolium hexafluorophosphate the excess volume is
positive~\cite{Henke2003}.

The excess enthalpy of the aqueous organic mixtures also show a
distinct  behavior. While the mixtures of water with small alcohol
molecules as
methanol~\cite{Marongiu1984,Randzio1986,Tomasziewicz1986} and
ethanol~\cite{Marongiu1984,Ott1986} exhibit a negative excess enthalpy,
the mixtures of water with large alcohol molecules as propanol and
butanol isomers show a positive excess enthalpy~\cite{Marongiu1984}.
Similarly to the small alcohol-water mixtures the excess enthalpy for
the water-alkanolamine solutions also show a
minimum~\cite{Mundhwa2007}.  In the case of  ionic liquids the excess
enthalpy also show two types of behavior. For the less hydrophobic
ionic liquids  in which the excess volume is negative, the excess
enthalpy is negative and shows a minimum at the same solute fraction
of the minimum of the excess volume. For the  hydrophobic ionic
liquids the excess enthalpy is positive and shows a maximum for the
same fraction of the solute of the maximum of the excess
volume~\cite{Miaja2009,Vatascin2015}.

The excess isobaric specific heat for the methanol at ambient
conditions increases with the fraction of the solute and exhibits a
maximum value around the solute concentration
$x_{2}=0.16$~\cite{Benson1980,Benson1982}. The excess free energy
presents a harmonic dependence on the methanol
fraction~\cite{Butler1933} and the excess entropy of mixing,
differently from the ideal  mixtures~\cite{McQuarrie1975}, assumes
negative values and decrease its value as the increasing methanol
concentrations~\cite{Lama1965}. In the case of the ionic liquids the
constant pressure heat capacity also shows an oscillatory behavior but
the peak occurs at higher concentrations of the solute
$x_{2}=0.3$~\cite{Miaja2009,Vatascin2015}.

The description of this complex behavior  of the organic solutes in
water in can be made, in principle,  in the framework of the
Frank and Evans~\cite{Evans1945} iceberg theory. These authors proposed
that water is able to form microscopic icebergs around solute
molecules depending on their size and the water-solute interactions.
Recent experiments~\cite{Dixit2002} using neutron diffraction support
Frank and Evans~\cite{Evans1945} scenario for the methanol. The
diffraction of a concentrated alcohol-water mixture ($x_{2} = 0.70$)
suggests that at these conditions  most of the water molecules
($\sim 87\%$) are organized in water clusters bridging  methanol
hydroxyl groups through hydrogen bonds. In the same direction an
experimental result from X-ray emission spectroscopy for an equimolar
mixture of methanol and water carried out by
Guo {\it et.al.}~\cite{Guo2003} suggests that in the mixture the
hydrogen bonding network of the pure components would persist to a
large extent, with some water molecules acting as bridges between
methanol chains.
Consistent with these results, recent experimental work for the
methanol~\cite{Dixit2002,Guo2003} suggests that the negative excess the
 entropy of mixing arises due to a relatively small degree of the
interconnection between the hydrogen bonding networks of the different
 components rather than from a water restructuring~\cite{Dixit2002}.

Motivated by these experimental results and by the huge number of
applications, water-methanol mixtures have been intensively studied by
computer simulations.
In these simulations, water molecules  are represented  by one of well
 known classical models SPC$/$E~\cite{Berendsen1987},
ST4~\cite{HeadGordon1993}, TIP5P~\cite{Jorgensen2000} and methanol
molecules are frequently modeled by OPLS force
field~\cite{Jorgensen1996}.
Using Molecular Dynamics simulation, Bako~{\it et.al.}~\cite{Bako2008}
found that on increasing the methanol fraction in the mixture, water
essentially maintains its tetrahedral structure, whereas  the number of
hydrogen-bonds is substantially reduced.
Allison~{\it et.al.}~\cite{Allison2005} showed that not only the number
hydrogen-bonds decreases, but the water molecules become eventually
distributed in rings and clusters in accordance with the experimental
results~\cite{Dixit2002}. Analyzing the spatial distribution function
of the water, Laaksonen~{\it et.al.}~\cite{Laaksonen1997} observed that
the system is highly structured around the  hydroxyl groups and that
the methanol molecules are solvated by water molecules, in accordance
with well known iceberg theory~\cite{Evans1945}.

In addition to the atomistic approaches,  water-methanol mixture has
been modeled by continuous potentials in which the water is
represented by a spherical symmetric two length scale potential while
the methanol is represented by a dimer in which the methyl group is
characterized by a hard sphere and the hydroxyl  is a water-like
group~\cite{hus2014,zhiqiang2012}. Numerical  simulations for this
system displays good qualitative agreement with the response functions
 for different temperatures~\cite{hus2014,zhiqiang2012} but fails to
produce the heat capacity behavior and does not provide the structural
 network observed in experiments and predicted by the iceberg  theory.

Due to the variety and complexity of the ionic liquids, very few
theoretical studies have been made for analyzing the ionic liquids
aqueous solutions. For example, there is no clear picture explaining
why the excess volume of some ionic liquids is negative while for
others is positive. In addition, it is not clear why for large
alcohols the excess enthalpy is positive while for the methanol is
negative. The explanation for these different behaviors both in the
alcohols and in the ionic liquids might rely in the disruption of the
iceberg theory as the solute is large of hydrophobic. In order to test
 this idea, here we explore how the excess properties of the
water-solute mixture is affected by the change of the water-solute
interaction from attractive to repulsive. In order to allow for the
water to form a structure not present in the continuous effective
potentials, our model exhibits a tetrahedral structure.

In this work the water and  the solute are  modeled following the
associating lattice gas model
(ALG)~\cite{Girardi2007,Szortyka2010,Szortyka2012} scheme. The two
molecules are specified by adapting the hydrogen bond and the
attractive interactions for each molecule. The excess volume and
enthalpy are computed for various types of water-solute interactions.

The remaining of the paper goes as follows. In the
Section~\ref{sec:model} the models for water, solute and mixture are
outlined  and the ground state behavior is presented. The technical
details about the calculations of ground state are presented in the
Appendix~\ref{ap:entropy}. In the Section~\ref{sec:methods} the
computational methods are described and the technical aspects can be
found in Appendix~\ref{sec.isop} and~\ref{sec.isopmix}. In the
Section~\ref{sec:results} results are presented.
Section~\ref{sec:conclusions} ends the paper with the conclusions.

\section{Model}\label{sec:model}

\begin{figure}[!htb]
  \includegraphics[clip,width=7cm]{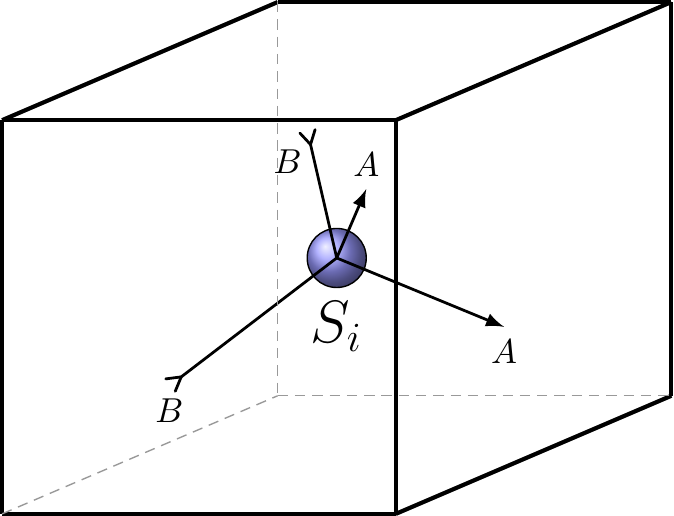}
\includegraphics[clip,width=7cm]{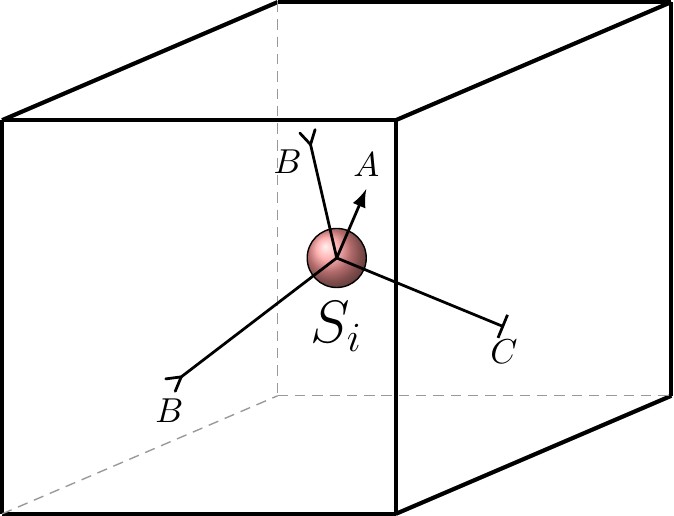}
    \caption{Representation of lattice for pure components. The blue
      sphere represents the water and the patches $A$ and $B$
      represent the acceptors and donors arms respectively. The red
      sphere represent the solute particle and the arms $A$ and $B$
      represent the acceptor and donors, and the patch $C$ represents
      the anisotropic group}
    \label{fig:model}
\end{figure}

We consider three systems: pure water, pure solute and water-solute
mixture. In the three cases the system is defined on a  body-centered
cubic (BCC) lattice.
Sites on the lattice can be either empty or  occupied by a water or by
a solute molecule.
Particles representing both water and solute molecules carry four
arms that point to four of the nearest neighbor (NN) sites on the BCC
lattice as illustrated by the figure~\ref{fig:model}.
The interactions between NN molecules are described in the
framework of the lattice patchy models~\cite{Almarza2012a}.
The particles carry eight patches (four of them corresponding
to the arms in the ALG model),
and each of the patches points to one of the NN sites in the BCC lattice as
illustrated in the figure~\ref{fig:model}.
The water molecules have two patches of the type $A$ (acceptors),
two patches of the type $B$(donors) and four
patches of the type $D$ (which do not participate in bonding interactions).
Since the patches of the types $A$
and $B$ participate in the hydrogen bonding, a water molecule can
participate in up to four hydrogen bonds.
The structure of the
solute is similar to the structure of
the  water, but it has only one patch of type $A$, the other patch $A$ is
replaced by a patch of the type $C$ that
represents the anisotropic group which makes
water and the solute different. In the case in which
the solute is
the methanol $C$ is the methyl group while
for other alcohols and ionic liquids
it does represent larger chains.

\begin{figure}[!htb]
  \includegraphics[clip,width=8cm]{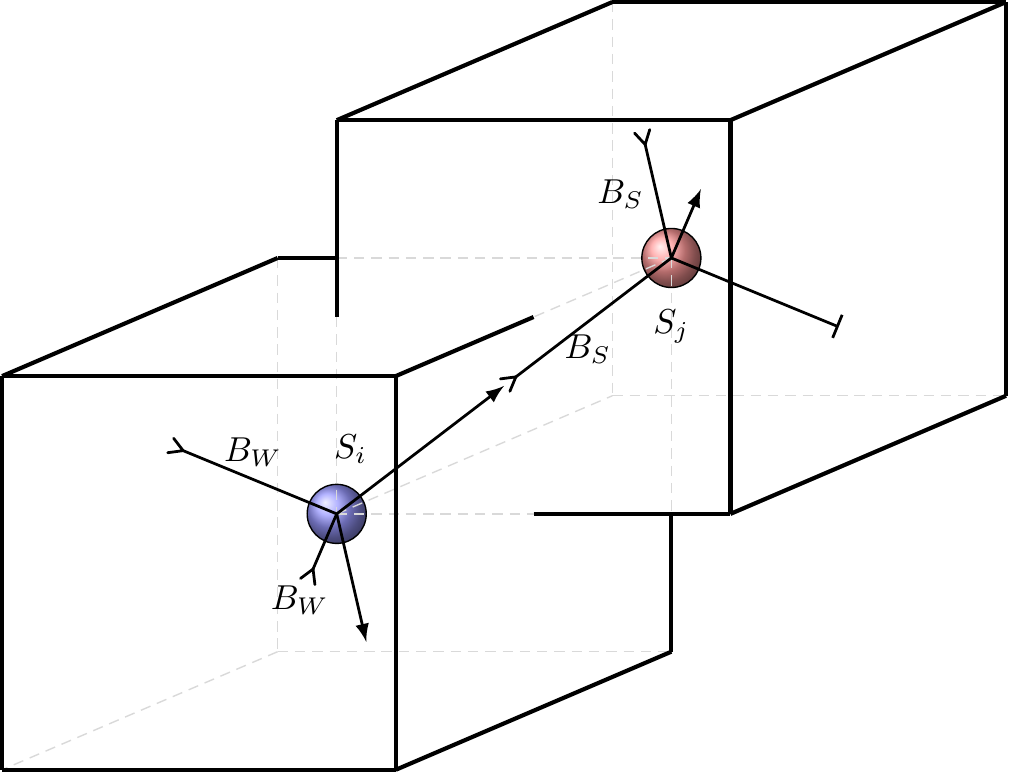}
  \caption{Representation of lattice. $S_i$ and $S_j$ represent
    particle on its respective positions $i$ and $j$. Blue sphere
    represent a water and red, a solute particle. $B_W$ and $B_S$
    represent the patches B of water and solute respectively. The
    patch $D$ is not represented here for the clarity of the image.}
  \label{fig:lattice}
\end{figure}
The distinction between patches implies $12$ possible orientations for
the water molecules and $24$ possible orientations for the  solute
molecules.

The potential energy is defined as a sum of interactions between
pairs of particles located at sites which are NN on the BCC lattice.
The interaction between particles $i$ and $j$, which are NN, only
depends on the type of patch of particle $i$ that points to particle
$j$, and on the type of patch of particle $j$ that points to particle
$i$.
The values of the interaction
as a function of the types of the two interacting patches
are summarized in the Table~\ref{tabla1}.
The interaction between occupied neighbor sites
is repulsive with an increase of
energy by $\epsilon_{ij}=\epsilon$ with the
exception of three cases.
For patch-patch interaction of type $A-B$
the energy interaction is taken as:
$\epsilon_{ij}=-\epsilon$.
If the interaction is of type $B-C$, with the $B$ patch
belonging to a solute molecule
there is also an attractive interaction
$\epsilon_{ij}=-\lambda_S\epsilon$ (with $\lambda_S>0)$, whereas if
the patch $B$ belongs to a water molecule
the interaction energy is given by
$\epsilon_{ij}=-\lambda_W\epsilon$.

We have considered $\lambda_S=0.25$, and three cases for the $B$-$C$
water-solute interaction: attraction with $\lambda_W = \lambda_S$,
non-interacting with $\lambda_W=0$ and repulsion with  $\lambda_W<0$.
The first case represents systems dominated by the water-solute
attraction. This is the case of the methanol in which it is assumed
that the methyl group shows a small but attractive interaction with
the water. This also represents the ionic liquids in which the anions
groups are hydrophilic and the cationic chains are not too
long~\cite{Freire2007,Ranke2009}. The second case represents alcohols
with larger non-polar alkyl substituents~\cite{Marongiu1984}. The
third case represents the ionic liquids in which the combination of
the anions and cations lead to an hydrophobic
interaction~\cite{Freire2007,Ranke2009}. Due to the simplicity of our
model solute, size and hydrophobicity effects are not taken into
account independently, but both are considered through the
$\lambda_W$ parameter.

\begin{table}
\begin{tabular}{|cc||rr|rr|rr|rr|rr|}
\hline
$p(S)$ && D &&  A && B$_{\rm W}$&& B$_{\rm S}$  &&  C & \\ \hline\hline
D &&  $\epsilon$ && $\epsilon$  &&  $\epsilon$ &&$\epsilon$  && $\epsilon$ & \\
A &&  $\epsilon$ && $\epsilon$  &&  $-\epsilon$ && $-\epsilon$ && $\epsilon$ & \\
B$_{\rm W}$ &&  $\epsilon$ && $-\epsilon$  && $\epsilon$ && $\epsilon$ &&
$-\lambda_{W}\epsilon$ & \\
B$_{\rm S}$ && $\epsilon$  && $-\epsilon$ && $\epsilon$  &&$\epsilon $ && $-\lambda_{S}\epsilon$  &\\
C && $\epsilon$  && $\epsilon$  && $-\lambda_W \epsilon$ && $-\lambda_{S}\epsilon$ && $\epsilon$ & \\ \hline
\end{tabular}
\caption{Interactions between NN particles of the same type (solute
  or water). The interaction depends on the patches of both particles
  involved in the interparticle bond. The interaction between patches
  of type C and B depends on the type of molecule: water (W) or
  solute (S) that provides the patch B. We consider
  $0 < \lambda_S \le 1$; and $\epsilon > 0$.
Patches of types A, B, and C correspond to the four arms of
the standard ALG model.
}
\label{tabla1}
\end{table}

At zero temperature for  the cases of the pure water and the pure
solute systems  three possible thermodynamic phases can appear in the
model: For low values of the chemical potential, $\mu$, the stable
phase is the empty lattice representing the gas phase at reduced
density, $\rho^* = N/N_L=0$, with $N$ being the number of particles
(occupied sites) and $N_L$ the number of sites of the lattice.
Increasing $\mu$ a low density liquid phase (LDL)
appears~\cite{Girardi2007,Buzano2008,Szortyka2010,Szortyka2012}, where
 half of the sites of the lattice are occupied by particles
($\rho^*=1/2$).
 These sites are those belonging to one of the diamond sublattices
\cite{Almarza2011b} that can be defined on the BCC lattice. Every
patch of the type $A$ is pointing to a patch of the  type $B$, and
{\rm vice versa}. In the case of water only
 pair interactions $AB$ occur. In the case of
the solute both $AB$ and $CB$ interactions occur.
At higher values of the chemical potential, the stable phase is the
high density liquid (HDL), where all the sites are occupied, and as
for the LDL phase, in the case of water, every patch of type $A$ is
bonded to a patch of type $B$ and {\em vice versa}. In the case of the
solute every patch of type $A$ is bonded to a patch of type $B$ and
every patch of type $C$ is bounded to a patch of type $B$.
The modification of the  ALG  model by considering different types of
arms introduce, at zero temperature, a residual entropy per particle
$s_0$, that in thermodynamic limit can be written as,
$s_0=k_B\lim_{N \to \infty} \left[ N^{-1} \ln{Q_0(N)} \right]$, where
$k_B$ is Boltzmann’s constant, and $Q_0(N)$ is the number of
configurations of the system, in which every patch of type B is
interacting with a patch of type A (or C), and every patch of type A
(or C) is interacting with a patch of type B. Using Monte Carlo (MC)
simulations and thermodynamic integration
techniques~\cite{AllenTildesleyBook,FrenkelSmitBook} we have obtained
the values of residual entropy for the  water
($s^{\rm (W)}_0 /k_B = 0.41041 \pm 0.00002$) and solute
($s_0^{\rm (S)} /k_B = 1.10356\pm 0.00002$) models. For more details
about the computation of residual entropies, see
Appendix~\ref{ap:entropy}.

From the values of residual entropy we can study the system in ground
state. The Grand Canonical thermodynamic potential can be written as:
\begin{equation}
\Phi \equiv -p V = U - T S - \mu N,
\end{equation}
where $U$ is the internal energy, $S$ is the entropy, and $N$ the
number of particles (occupied positions). In the Ground State, the
stable phase for a given value of $\mu$ is the one with the  minimum
value of $\Phi$.
 Considering the description of the ordered phases explained above,
for the water model $\Phi$ take the values (for $T\rightarrow 0$):
\begin{equation}
\begin{array}{llllr}
\Phi^{\rm (W)}_G(V,\mu)/V^* & = & 0 & & (\rho^*=0); \\
 \Phi^{\rm (W)}_{LDL} (V,\mu) /V^* & = &
- \epsilon -Ts^{\rm (W)}_0 /2 - \mu/2;
 & & (\rho^*=1/2);\\
\Phi^{\rm (W)}_{HDL} (V,\mu)/V^* & = & -  T s^{\rm (W)}_0 - \mu; & &
(\rho^*=1)
\end{array}
\label{eq.gsw}
\end{equation}
with $V^*$ being the reduced volume (equal to the number of sites).
Imposing that in coexistence the
$\Phi^{\rm (W)}_G=\Phi^{\rm (W)}_{LDL}$ and
$\Phi^{\rm (W)}_{LDL}=\Phi^{\rm (W)}_{HDL}$ we obtain the values of
the chemical potential and the pressure, at the transitions in
the limit of low temperatures
\begin{equation}
\begin{array}{lll}
\mu = - 2 \epsilon - T s_0^{\textrm{(W)}}, & \quad
 p w_0/\epsilon = 0, & \quad
  \textrm{G-LDL water}\\
\mu =  2 \epsilon - T s_0^{\textrm{(W)}}, & \quad
p w_0/\epsilon = 2, & \quad
\textrm{LDL-HDL water};
\end{array}
\label{eq.coexw}
\end{equation}
where the factor $w_0 = V/N_L$ correspond to the volume per site.
For the case of pure solute the thermodynamic potential
$\Phi$ of the different phases
as $T\rightarrow 0$ is given by
\begin{equation}
\begin{array}{llllr}
\Phi_G^{\rm (S)} (V, \mu) /V^*    & = & 0  ; &  \ \ & ( \rho=0); \\
\Phi_{LDL}^{\rm (S)}(V,\mu) /V^* & = & -(1+\lambda_S) \epsilon /2 -
T s_0^{\textrm{(S)} }/2  - \mu/2; & &  (\rho=1/2) ;\\
\Phi_{HDL}^{\rm (S)}(V,\mu) /V^*  & = & (1-\lambda_S) \epsilon  -
T s_0^{\textrm{(S)} }
- \mu; & & (\rho=1),
\end{array}
\label{eq.gsm}
\end{equation}
and for the  phase equilibria at low temperature we get:
\begin{equation}
\begin{array}{lllll}
\mu  = - (1+\lambda_S) \epsilon - T s_0^{\textrm{(S)}}, &&
p w_0/\epsilon=0,  &&  \textrm{G-LDL solute}, \\
\mu =  (3-\lambda_S) \epsilon- Ts_0^{\textrm{(S)}},  &&
p w_0/\epsilon=2, && \textrm{LDL-HDL solute}
\end{array}
\label{eq.coexm}
\end{equation}
All the relevant quantities will be expressed in reduced units, such
as:
\begin{equation}
  \label{eq:red_units}
  \mu^*=\dfrac{\mu}{\epsilon} \quad , \quad
  T^*=\dfrac{k_BT}{\epsilon} \quad , \quad
  c_{V}^*=\dfrac{c_{V}}{k_B}, \quad
  p^*=\dfrac{p w_0}{\epsilon}, \quad
  s^*=\dfrac{s}{k_B}
\end{equation}

\section{Simulation and Numerical Details}
\label{sec:methods}

In order to obtain the phase diagrams and compute the thermodynamic
and structural properties of one-component systems, we have performed
MC simulations in the grand canonical ensemble (GCE) for system sizes
  $512 \le N_L \le  65536 $ where $N_L=2 L^3$. The simulations have
used $\sim  8 \times 10^6$ MC sweeps for equilibration  and
$\sim 4 \times 10^6$ sweeps  for evaluating the relevant quantities.
Each MC sweep is defined as $N_L$ one-site attempts to generate a new
configuration.
Each attempt is carried out as follows: i) A site  $i$ on the lattice
is chosen at random;
this site can adopt $n_s$ possible states, $S_i=0,1,2 \cdots, n_s-1$
($n_s=13$ for pure water; $n_s=25$ for pure solute, and $n_s=37$ for
 the mixtures);
$S_i=0$ represents an empty site, and the remaining values stand for
the different species that can occupy the site and their respective
orientations.
ii) For the selected site, one of its possible $n_s$ states
is selected at random, with probabilities given by:
%
\begin{equation}
P(S_i) \propto   \exp \cch{ -\frac{ U_i(S_i) - \mu(S_i)}{k_B T}},
\hspace{1cm} S_i=0,1,2, \cdots, n_s;
\end{equation}
where $U_i(S_i)$ contains the potential energy interactions between
site $i$ at state $S_i$ with its NN, and $\mu(S_i)$ is the chemical
potential of the component associated with state $S_i$. Notice that
for an empty site $S_i=0$, both $U_i(S_i)$ and $\mu(S_i)$ are zero,
and that for $S_i \ne 0$, the value of $U_i$ depends on the states of
the sites which are NN of $i$.

We have combined the one-site sampling procedure  with different
advanced techniques in order to enhance the simulation efficiency:
In the regions close to continuous transitions, where critical slowing
 down may be present~\cite{LandauBinderBook}, we have made use of the
  Parallel Tempering (PT) method\cite{Swendsen1986a,Earl2005a}.
The Gibbs-Duhem integration~\cite{Kofke1993} technique adapted for
working in the GCE~\cite{Almarza2011b,Almarza2012a} was employed in
the location of the discontinuous phase transitions of the system.
In order to study the excess properties of the mixtures as functions
of the pressure and temperature we have developed methods to build
up isobars for one-component systems (See Appendix~\ref{sec.isop}),
and lines at constant pressure and temperature with varying composition
for the binary mixtures (See Appendix~\ref{sec.isopmix}).

\section{Numerical Results}
\label{sec:results}

\subsection{The phase diagram for pure components}

\begin{figure}
  \includegraphics[clip,width=8cm]{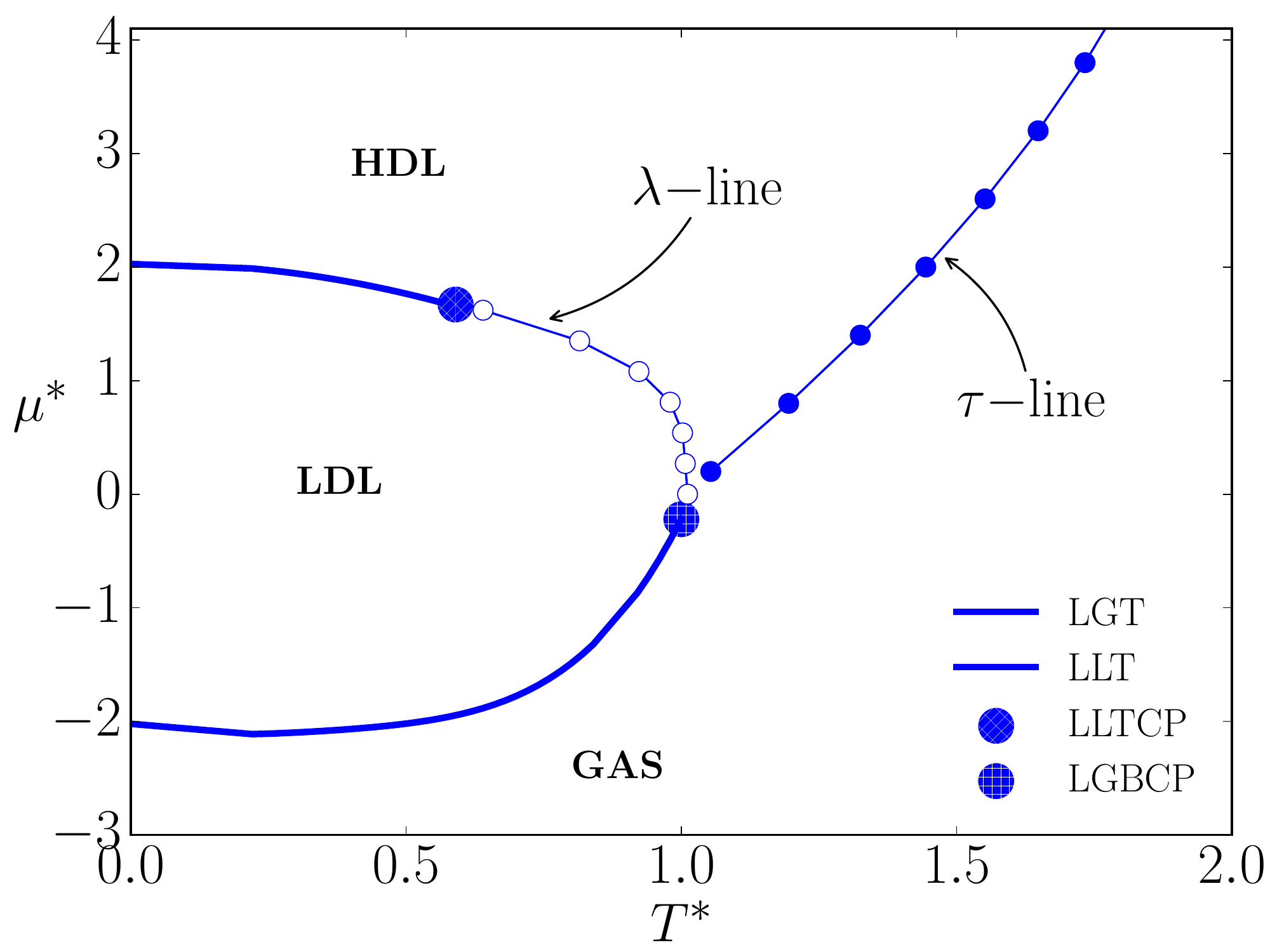}
  \includegraphics[clip,width=8cm]{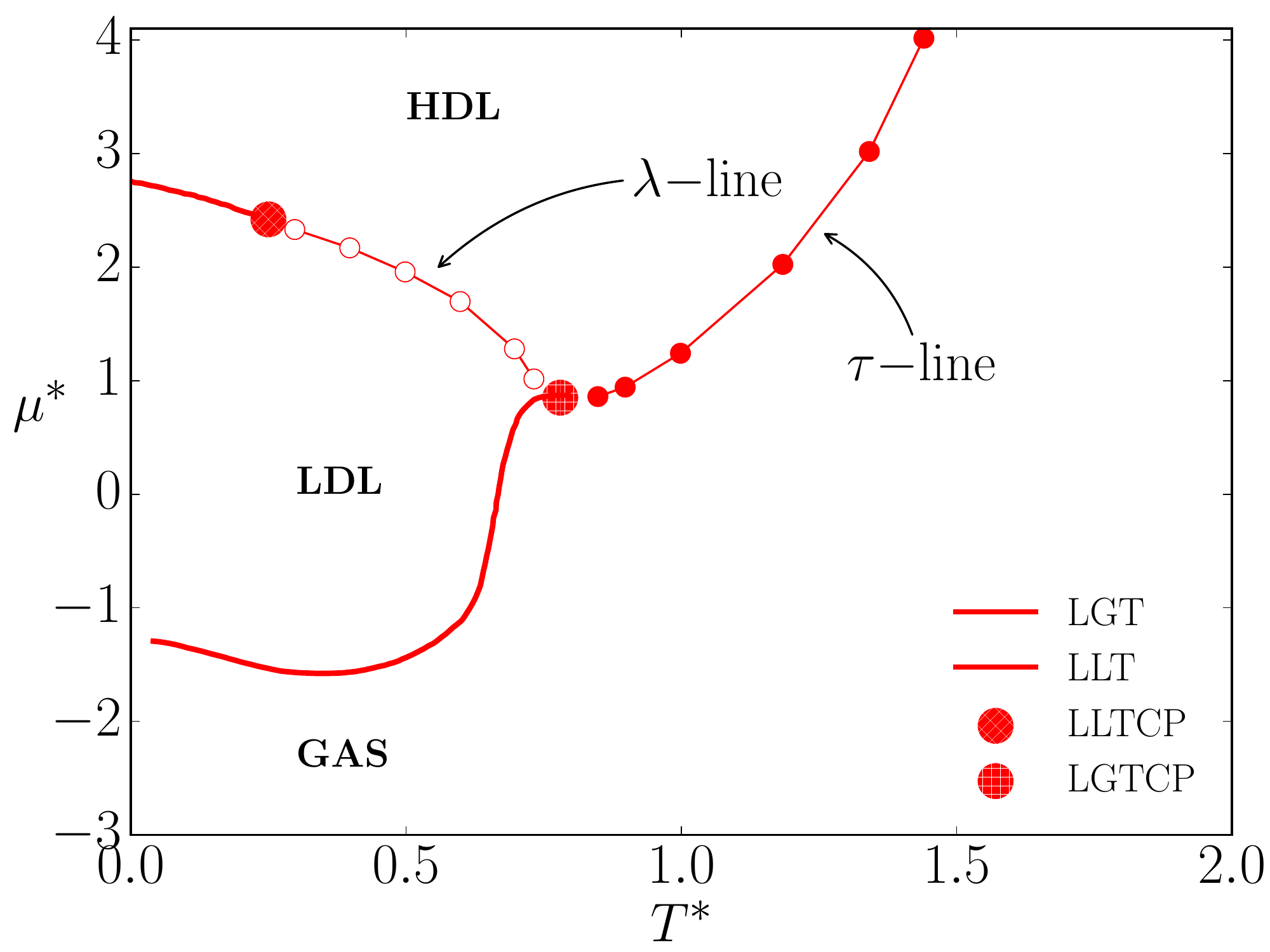}
  \caption{Reduced chemical potential versus reduced
    temperature phase diagram for (a) pure water  (b) pure solute.
    Solid thick lines represent first-order phase transitions,
    liquid-gas (LGT) and liquid-liquid (LLT). Thin lines represent
    continuous (second-order) transition, the  empty circles show the
    $\lambda-$line and the filled circles are the $\tau-$line. The big
    patterned circles represent the multi-critical points of the model.
    (See the text for details).}
  \label{fig:WT_muxT}
\end{figure}

The chemical potential $vs$. temperature phase diagrams are shown in
the Fig. \ref{fig:WT_muxT} for the pure water and  for the pure
solute. Three different phases, G, LDL, and HDL appear, as expected.
At low temperature, there are two, G-LDL and LDL-HDL first-order
transitions. The first order LDL-HDL transition finishes, both for
water and solute, in a liquid-liquid tricritical point (LLTCP). The
LLTCPs occur at $T^*_{tc}\simeq 0.59$ and  $\mu^*_{tc}\simeq 1.67$ for
water and at  $T^*_{tc}=0.25$ $\mu^*_{tc}=2.42$ the solute. Above
$T_{tc}$ the LDL-HDL transition becomes continuous and defines the
$\lambda-$line (See Fig. \ref{fig:WT_muxT}).
At high temperature, it appears a continuous transition between G and
HDL phases ($\tau-$line in Fig. \ref{fig:WT_muxT}).

The coexistence line between the G and LDL phases for water extends up
 to a bicritical point (LGBCP)~\cite{LavisBellBook} located at
$T_{bc}^*\simeq 1.00$ and $\mu_{bc}^* \simeq -0.22$. At this LGBCP the
G-LDL transition meets the lines for the critical G-HDL ($\tau-$line)
and LDL-HDL ($\lambda-$line) transitions.
In the case of solute, the G-LDL first order transition meets the
$\lambda$-line at an end point located at $T_{t}^* \simeq 0.74 $,
$\mu_{t}^* \simeq 0.86$. Above this temperature there is a G-HDL first
 order transition up to a tricritical point (LGTCP) located at
$T_{tc}'^* \simeq 0.85$, $\mu_{tc}'^* = 0.84$. Above this temperature
the G-HDL transition becomes continues and defines the $\tau$-line.

The continuous $\tau$ and $\lambda$ transitions, illustrated in the
figure~\ref{fig:WT_muxT}  are represented by thin lines and circles.
The values of the temperatures and chemical potentials for the
critical lines were obtained by computing appropriated order
parameters and their associated moments or cumulants.

\begin{figure}[!htb]
  \includegraphics[clip,width=8cm]{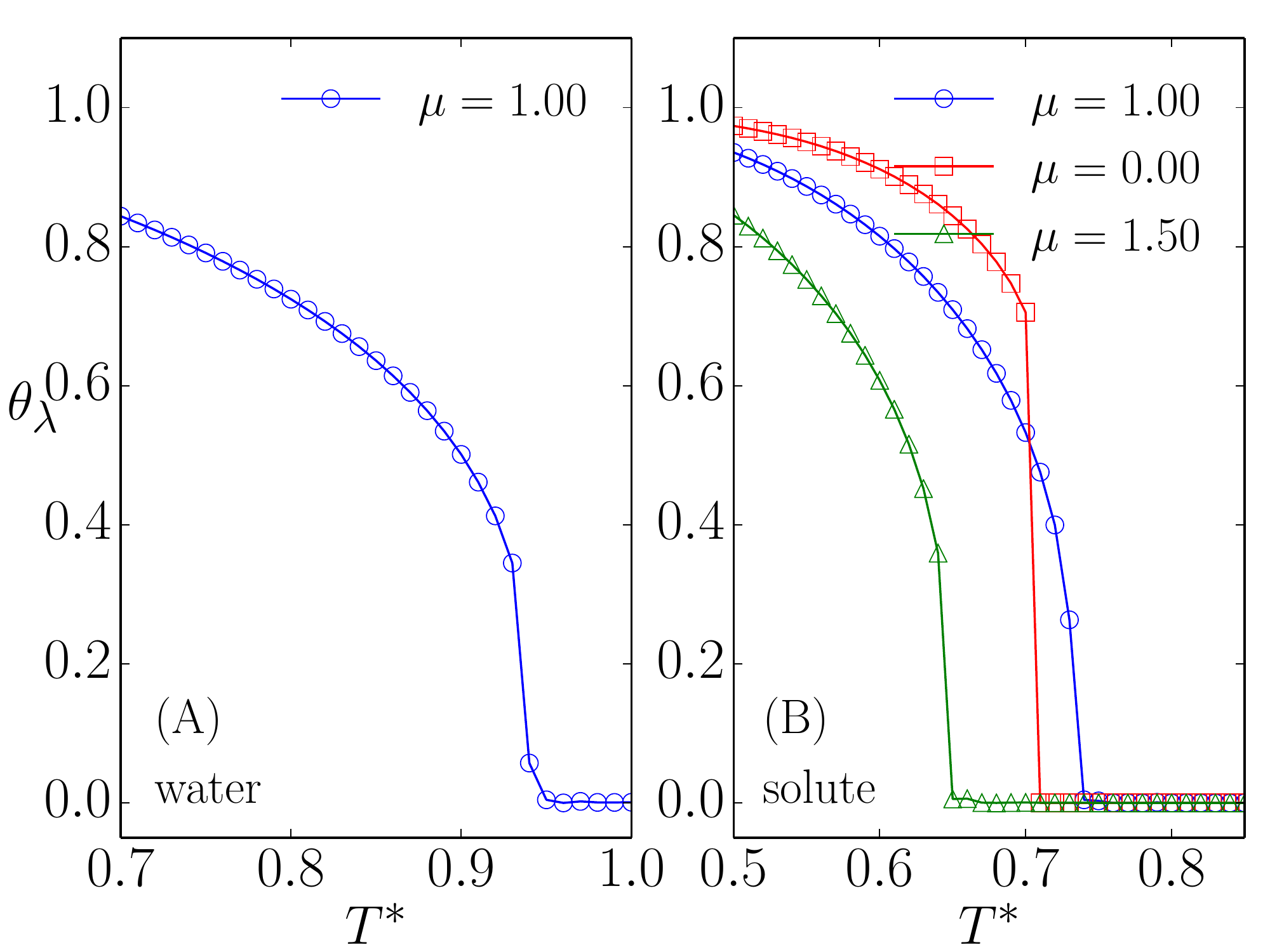}
\includegraphics[clip,width=8cm]{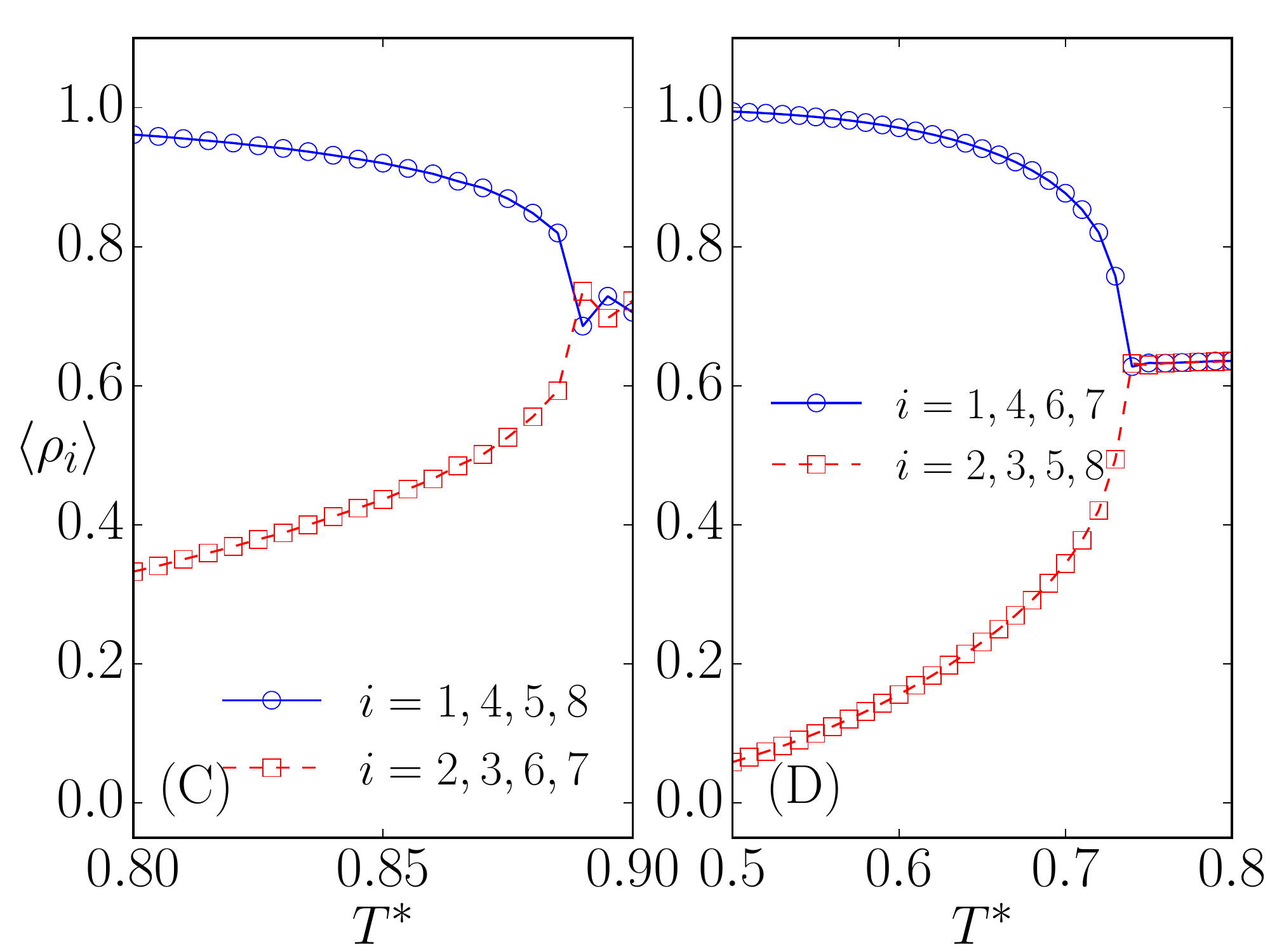}
  \caption{Order parameter  $\theta_t$ versus reduced temperature for
    (A) water and (B) solute for various chemical potentials. Average
    density of the empty (red) and full (blue) sublattices for (C)
    water and for (D) the solute.}
  \label{fig:thetat_param}
\end{figure}

In the case of the  $\lambda$ line, the $\theta_{\lambda}$ order
parameter is defined as follows. The system is divided into eight
sublattices~\cite{Szortyka2010}. The figure~\ref{fig:thetat_param}
illustrates the behavior of the eight sublattices as a function of
 the temperature at the $\lambda$-line. As the temperature is
decreased four sublattices become full while other four stay empty.
Then, from the density of these sublattices, the order parameter is
defined by
\begin{equation}
  \theta_{\lambda}=\dfrac{2}{V}\cch{\sum_{i=1}^{\rm full}\rho_i-
    \sum_{j=1}^{\rm empty}\rho_j},
\label{eq:op}
\end{equation}
where the index $i$ runs over the four sublattices which become full
at the LDL phase while the subindex $j$ runs over the four sublattices
 which remain empty at the LDL phase. The
figure~\ref{fig:thetat_param}  illustrates the value of this order
parameter as a function of the temperature for fixed chemical
potentials for both the pure water and the pure solute cases showing
the transition from all the sublattices equally populated to a
preferential occupation in four sublattices.

\begin{figure}[!htb]
   \includegraphics[clip,width=8cm]{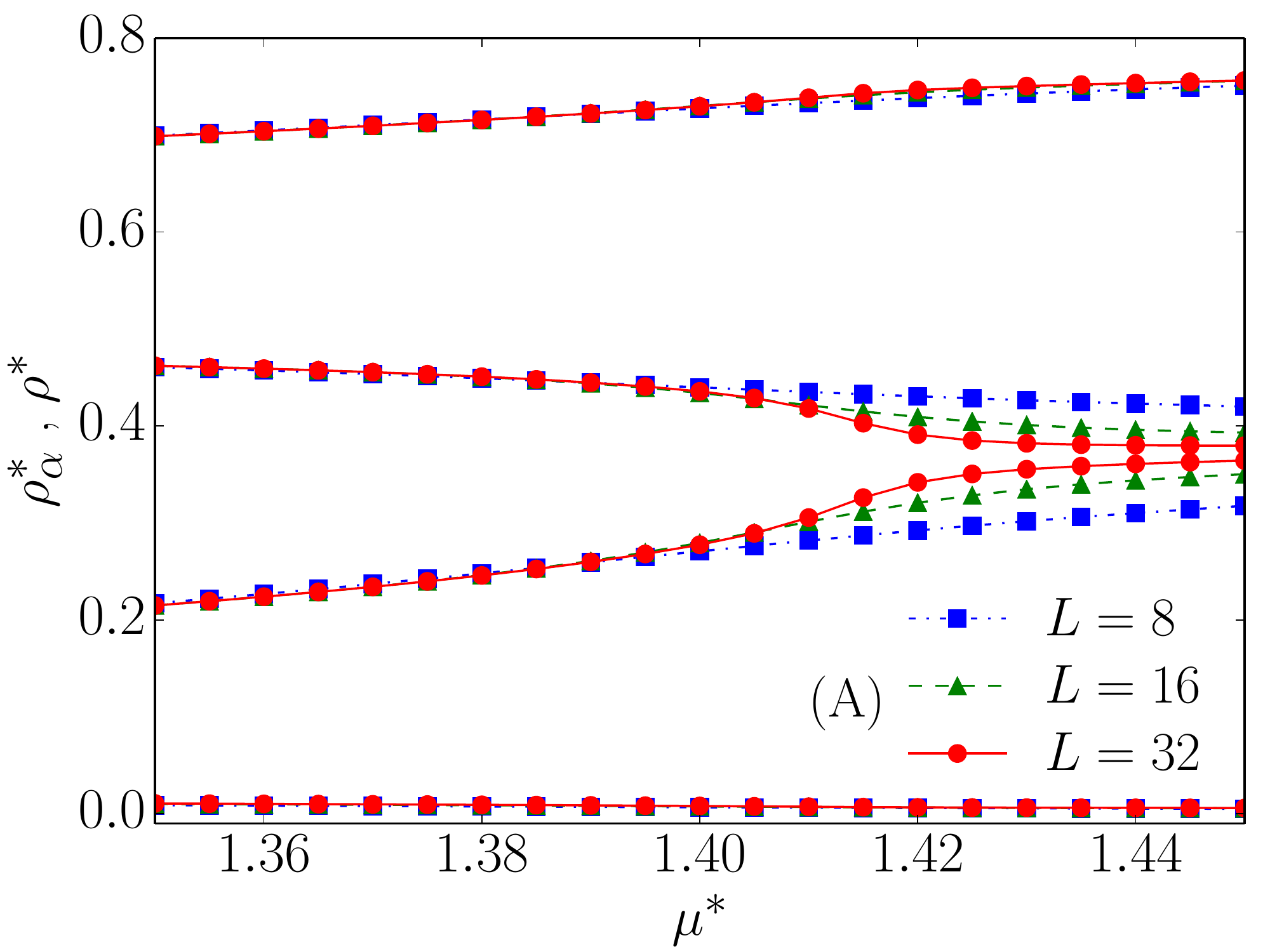}
   \includegraphics[clip,width=8cm]{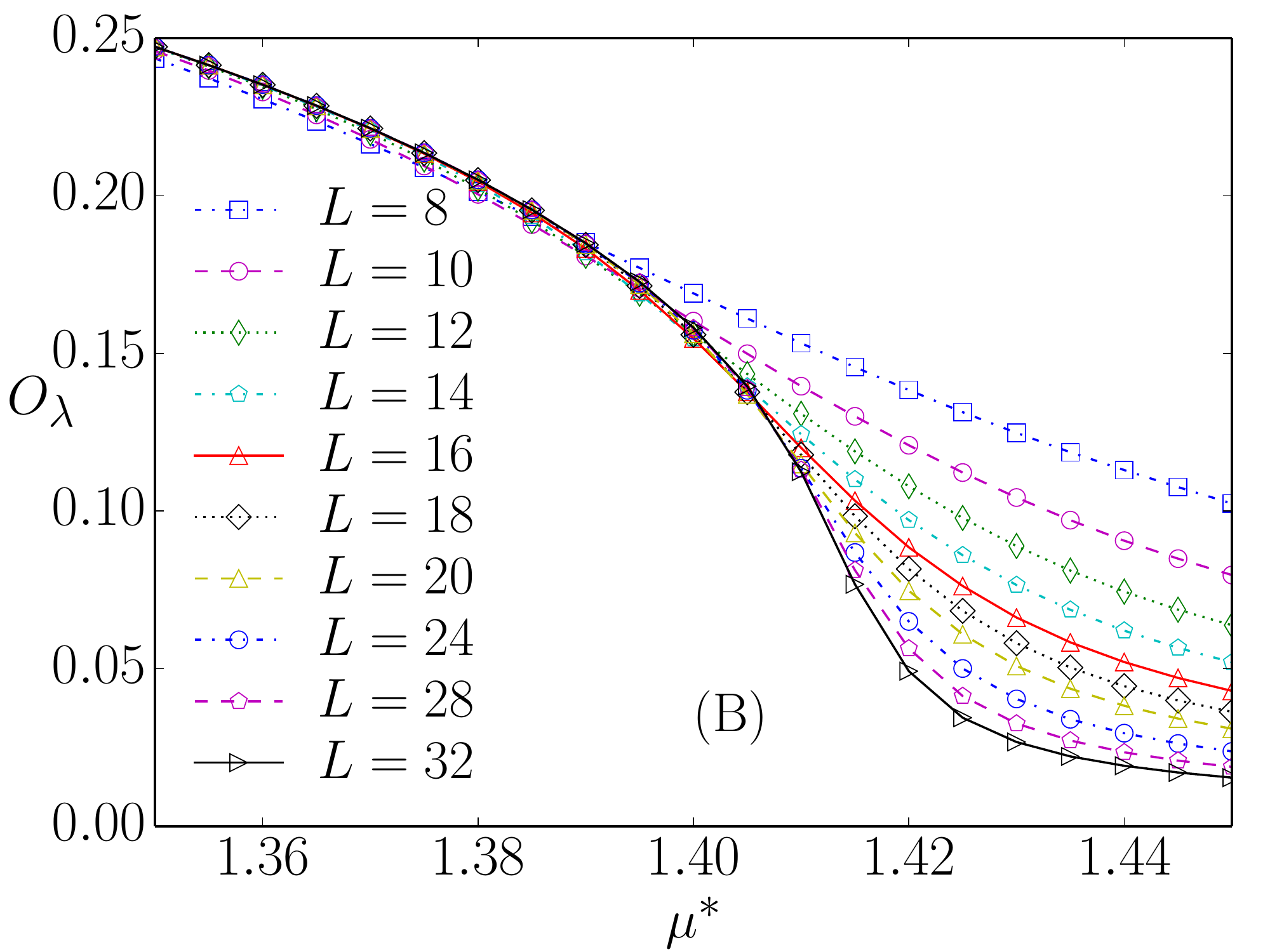}
   \includegraphics[clip,width=8cm]{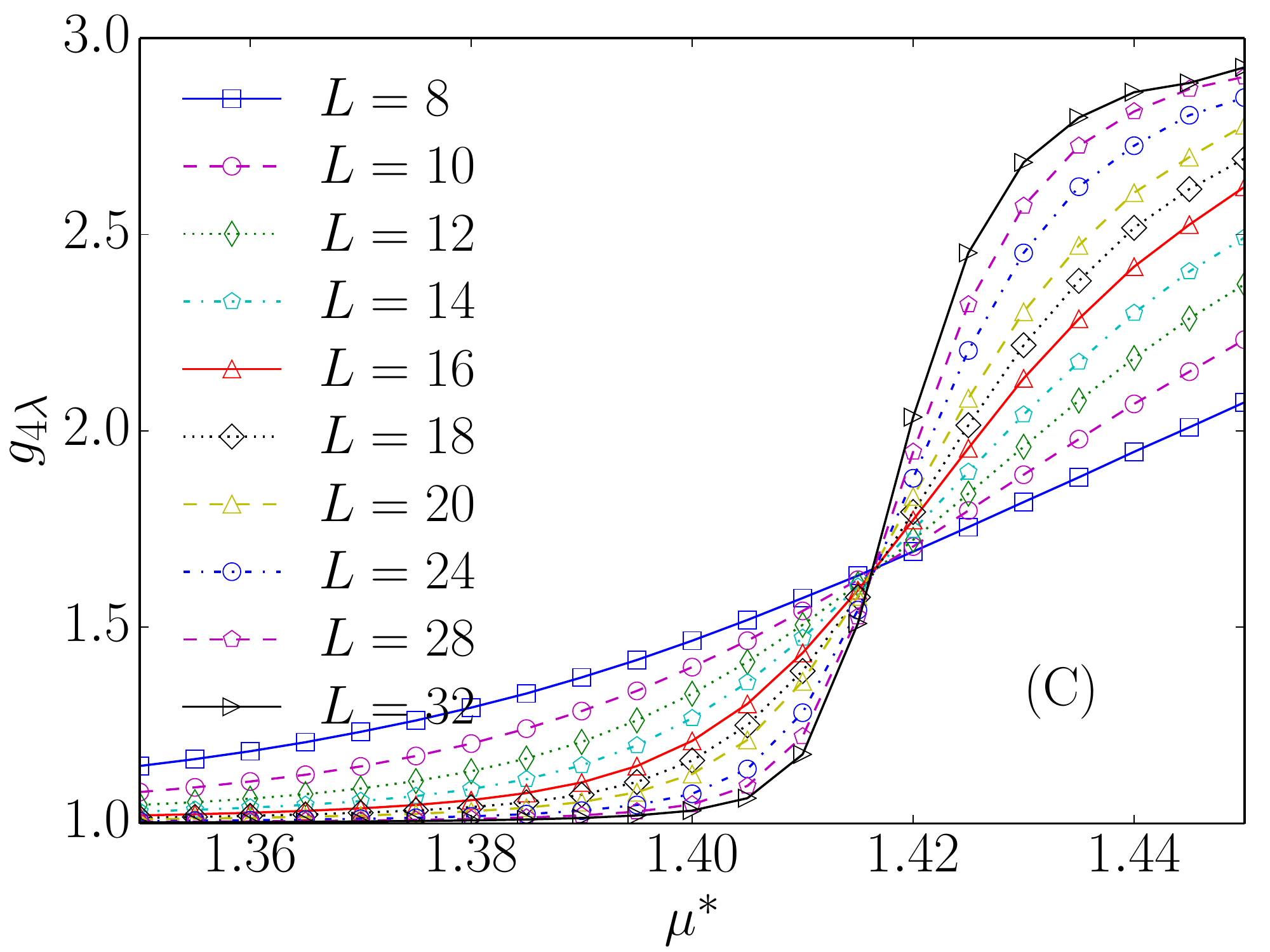}
  \caption{Location of the continuous LDL-HDL transitions for the
    water model at $T^*=0.80$ and different system sizes:
    $N_L = 2 L^3$. At (A) Total reduced density and partial
    densities, $\rho_{\alpha}$  as functions of $\mu$. (B) (left) Order
    parameter $O_{\lambda}$ and (left) $g_{4\lambda}$  as a
    function of $\mu$. }
  \label{fig:od}
\end{figure}

This transition can also be described by taking into account that for
the ALG model there are four possible realizations of the LDL
structure (with diamond structure) on the BCC lattice. The occupancy
of the sites and orientations of the arms (patches of types A, B and
C) is well defined for each LDL realization. Taking into account the
orientation of the occupied sites on a given configuration, we can
compute the number of particles in the system compatible with each of
the four LDL realizations. Let $n_i$, with $i=1,2,3,4$ be those
numbers. From each configuration, we can sort the $n_i$ values so that
$n_a \ge n_b \ge n_c \ge n_d$, and compute their corresponding densitie
s $\rho_a=n_a/N_L, \rho_b = n_b/N_L \cdots$. In the thermodynamic limit
 ($N_L \rightarrow \infty$), we expect for the G phase:
$\rho_a \simeq \rho_b \simeq \rho_c \simeq \rho_d \simeq \rho/4$. For
 the HDL phase $\rho_a \simeq \rho_b \gg \rho_c \simeq \rho_d$, and
finally for the LDL phase $\rho_a \gg \rho_b$. Accordingly the
presence of the LDL phase can be detected by an order parameter,
$O_{\lambda}$ ,given by
\begin{equation}
O_{\lambda} = \rho_a - \rho_b;
\end{equation}
The system size dependence of the shape the $O_{\lambda}$ distribution
can be analyzed by looking at the ratio~\cite{LandauBinderBook}:
\begin{equation}
g_{4\lambda} = \frac{ \aver {O_{\lambda}^4}}{\aver{O_{\lambda}^2}^2};
\label{eq.g4d}
\end{equation}
where the angular brackets represent average values.
In figure~\ref{fig:od}, we show the results for
$\langle O_{\lambda} \rangle$  and $g_{4\lambda}$ as functions of the
chemical potential for various lattice sizes and at $T^*=0.80$. The
crossing of the lines of $g_{4\lambda}$ for different values of $L$
locate the critical chemical potential at that temperature, i.e, the
corresponding point of the $\lambda-$line.

Complementary to the study of the $\theta_{\lambda}$ and $O_{\lambda}$
order parameters described above, the behavior of the  specific heat
at constant volume for different system sizes were analyzed.

At criticality, it is expected that the  specific heat  would show
a divergence as the thermodynamic limit is approached. The finite-size
 scaling behavior of the critical exponent of the specific heat,
$\alpha$, gives the critical behavior at the infinite
system~\cite{LandauBinderBook}. The heat capacity at constant volume
( per lattice site)
$c_V= \left( \partial {U}/\partial T \right)_{N,V}/V$ is computed from
the data obtained from simulations at constant chemical potential
through the expression~\cite{AllenTildesleyBook},
\begin{equation}
  c_{V}=\dfrac{1}{k_BT^{2}V}\cch{\aver{\delta{U}^2}
    -\dfrac{\aver{\delta {U}\delta N}^2}{\aver{\delta N^2} }}\; .
  \label{eq:spec_heat}
\end{equation}

Here $U$ is the interaction energy of model described in the
Table~\ref{tabla1}, and $N$ is the number of particles. The averages
in Eq.~(\ref{eq:spec_heat}) are carried out on the grand canonical
ensemble.
\begin{figure}[!htb]
  \centering
    \includegraphics[clip,width=8cm]{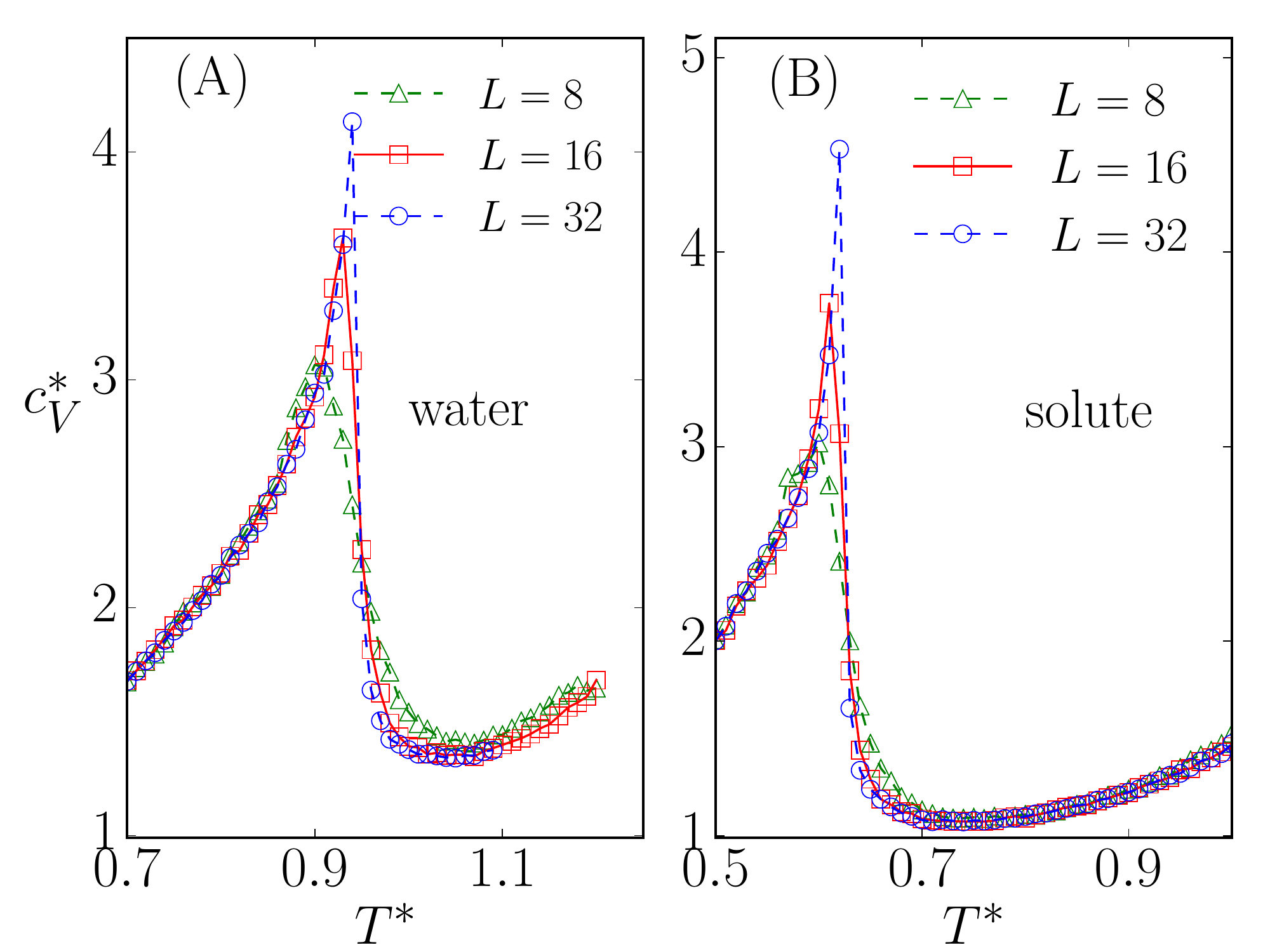}
    \includegraphics[clip,width=8cm]{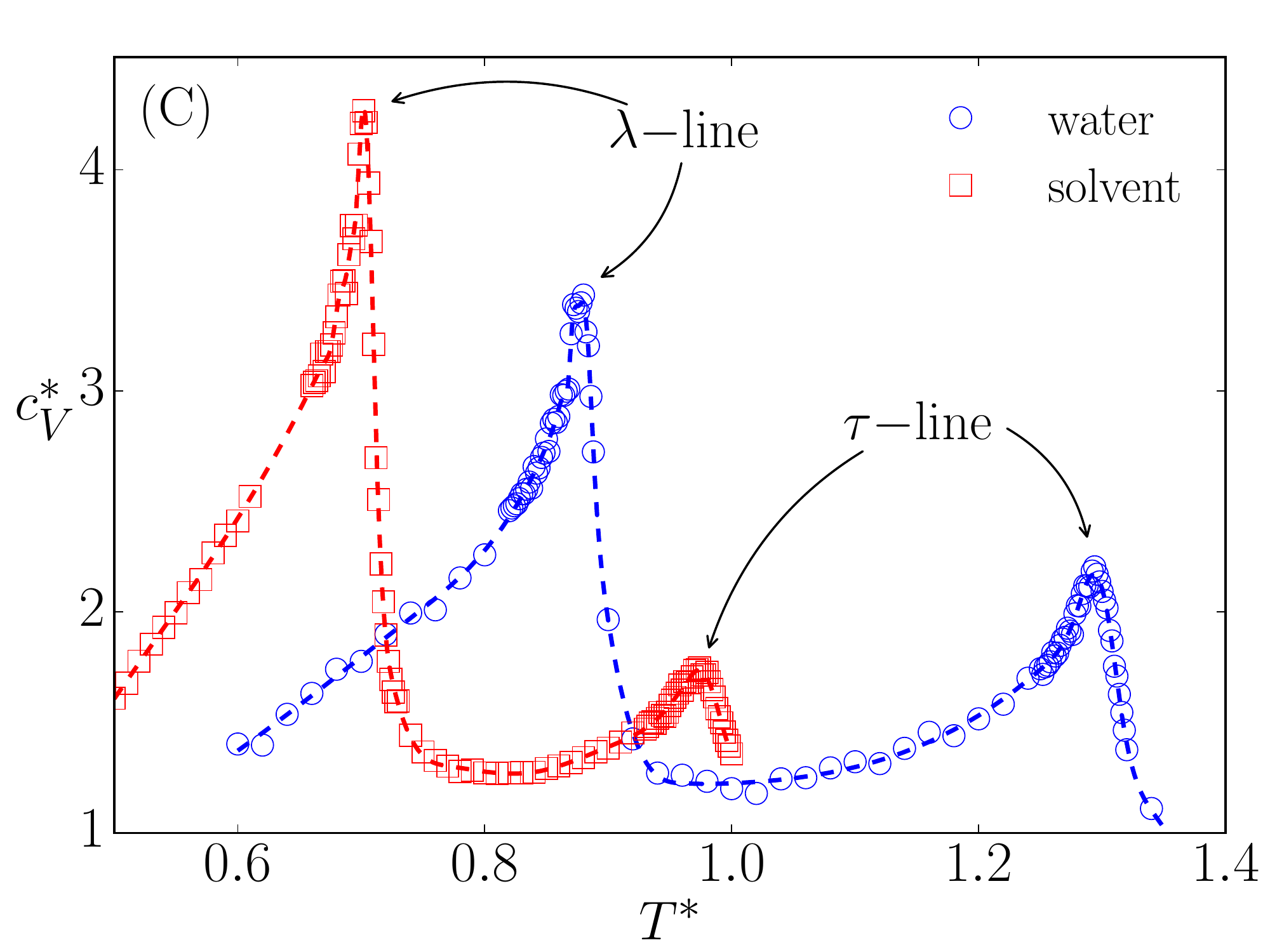}
  \caption{Heat capacity versus reduced temperature for the reduced
    chemical potentials (A) $\mu^*=1.0$ and  (B)$\mu^*=1.6$ for water
    and solute respectively for different $L$ values for the
    $\lambda$-line. The same for $\mu^*=1.2$ showing both the
    $\lambda$ and the $\tau$-lines. }
  \label{fig:cvxT}
\end{figure}

The figure~\ref{fig:cvxT} shows the specific heat at constant volume
{\em versus} temperature at constant $\mu$, illustrating the diverging
 peak at $T^*\simeq  0.9$ for the pure water system and at
$T^*\simeq 0.6$ for the pure solute, as $L$ increases. The peak in the
 heat capacity $c_V$ in addition to the $O_{\lambda}$ and
$\theta_{\lambda}$ behavior is employed to locate the $\lambda-$line.

\begin{figure}[!htb]
\includegraphics[clip,width=8cm]{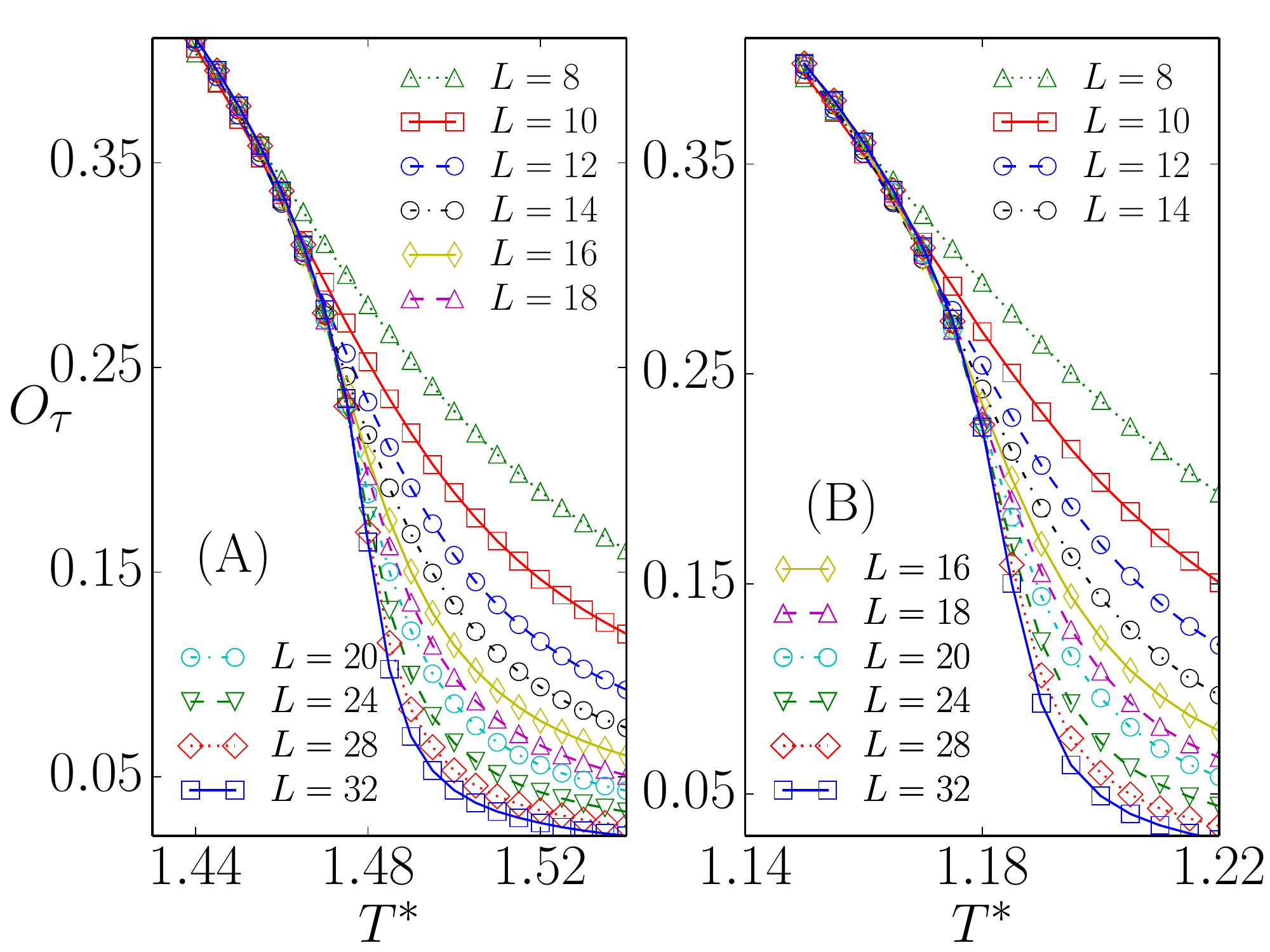}
\includegraphics[clip,width=8cm]{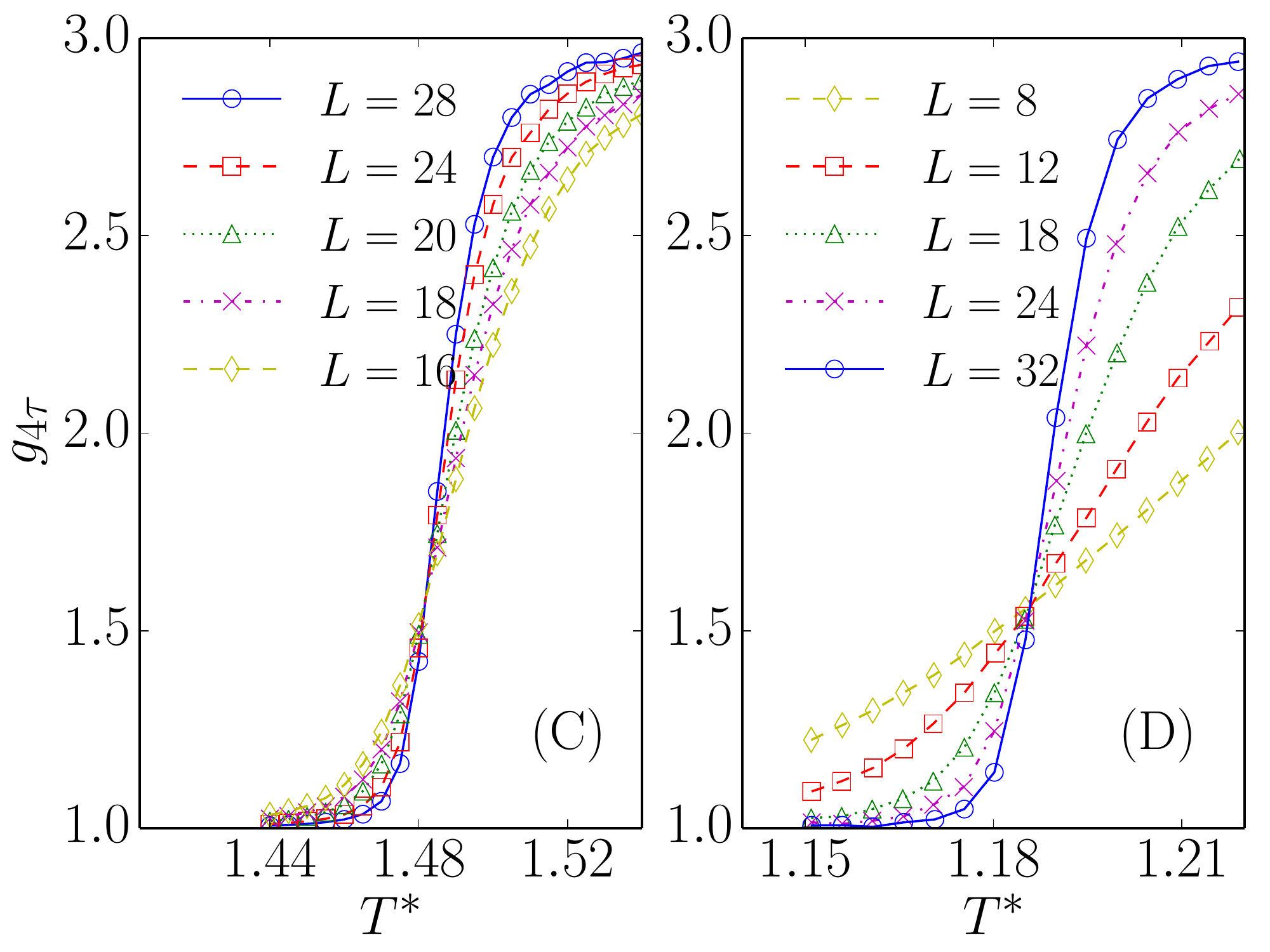}
\caption{$O_{\tau}$  versus reduced temperature for different systems
sizes at chemical potential $\mu=2.0$ for water (A) and solute (B)
models. $g_{4\tau}$ versus reduced temperature for different systems
sizes at chemical potential $\mu=2.0$ for water (C) and solute (D)
models.}
  \label{fig:g4xT}
\end{figure}

The $\tau$-line corresponds to the transition between G and HDL
phases. An order parameter based on the symmetry of the ALG model can
be defined to quantify the HDL ordering of the configurations. The BCC
 lattice can be splitted into two interpenetrated cubic sublattices.
In the HDL structure, each sublattice adopts a different but
complementary orientation. The appropriate order parameter for the
$\tau$-line is given by
\begin{equation}
  O_{\tau} = \frac{1}{N_L} \left| \sum_{i=1}^{N_L}  l(i) s(i) \right|,
\end{equation}
where $l(i)$ depends on the cubic sublattice where the site $i$ stands
 (with values $-1$, and $+1$ for the two sublattices), and $s(i)$
represents the orientation of the particle at site $i$, with $s(i)=0$
for empty sites.
This order parameter is analogous to that for antiferromagnetic Ising
models in bipartite lattices.
In the figure~\ref{fig:g4xT} we show the shape of the order parameter
at the HDL-G transition (top panels).
For the G phase $O_{\tau}$ vanishes in the thermodynamic limit while for
the HDL it remains finite. The approximate location of the transition
is obtained by the crossing of the curves for different sizes.
Even though the behavior of the order parameter illustrates how the
structure of the phases change at the phase transition, it does not
provide the precise temperature and chemical potential. The precise
location of the transitions can be achieved by looking at the system
size dependence of the ratio, $g_{4\tau}$:
\begin{equation}
g_{4\tau} = \frac{ \aver{O_{\tau}^4} }{\aver{O_{\tau}^2}^2 },
\end{equation}
where the brackets indicate average over grand canonical simulations.
At the $\tau$-line transition, it is expected that the values of
$g_{4\tau}$ become independent of the system size. The
figure~\ref{fig:g4xT}, examples for water and solute of the behavior
of $g_{4\tau}(T,N_L)$ for fixed $\mu$, at the $G-HDL$ transition
($\lambda-$ line) are presented. The value of $g_{4\tau}$ at the
crossing region, together with the form of the order parameter suggest
 three-dimensional Ising criticality. In addition the location of the
$\tau$-line can be confirmed by the divergence of the heat capacity.

\subsection{The excess properties of the water-solute mixtures}

Next, we explore the mixture of water and solute. The thermodynamic
excess properties of the mixture are defined by comparing the values
of a given extensive property per mol (or per molecule) with the
values of this quantity for an ideal mixture. In the case of the
excess volume we have:
\begin{equation}
{\bar V}^E(x,p,T) = {\bar V}(x,p,T) -\left[  (1-x) { \bar V}_1^0(p,T)
+ x {\bar V}_2^0(p,T)\right];
\label{eq:def_exc}
\end{equation}
where ${\bar V}(x,p,T)$ is the volume per molecule of the mixture at
molecular fraction $x$ of the solute (component 2) $x$
(i.e. $x\equiv x_2$), ${\bar V}_1^0(p,T)$ and ${\bar V}_2^0(p,T)$ are
the volumes per molecule  of the pure solvent (component $1$) and pure
 solute respectively.

The thermodynamic properties for different compositions at constant
$T$ and $p$ were computed using GCE simulations coupled to the
integration schemes explained in Appendices~\ref{sec.isop}
and~\ref{sec.isopmix}. In practice, for one component systems we apply
 an integration scheme to find  the line, $\mu(T|p)$ in the plane
$\mu-T$ that corresponds to a fixed value, $p$ of the pressure, and
for the mixtures we calculate the line $\mu_1(\mu_2|T,p)$ in the plane
 $\mu_1-\mu_2$ that keep fixed the values of $T$ and $p$.

The volumes per molecule was estimated from the simulations as:
${\bar V} = V/\langle N \rangle$.
The enthalpy, $H$ of a given system is given by: $H = U + p V$, where
$U$ is the internal energy (given for the patch-patch interactions).
The enthalpy per molecule can be estimated as:
$\bar{H} = \left[ \langle U \rangle + p V \right]\langle N \rangle$.
%
Whereas the mole fraction for given values of the activities
$z_i = \exp \left[ - \mu_i/(k_B T) \right]$, is computed as:
$\langle x \rangle \simeq \langle N_2 \rangle / \langle N \rangle$.

The integration procedure provides the results for the different
properties at equally spaced discrete values of the activity, $z$, of
one of the components, (say component 1) which span from
$z_1= z_1^{(0)}$ (pure solvent) to $z_1=0$ (pure solute). For each of
these cases the properties of interest, $x(z_1)$, ${\bar V} (z_1)$,
${\bar H}(z_1), \cdots$ are computed. Then, to estimate the dependence
 of the molar properties with the composition these properties are
fitted to polynomials of $x$ as:
\begin{equation}
\bar{Y}_f(x,T,p) = \sum_{j=0}^{j_{max}} a_{j}^{(Y)}(T,p)  x^j.
\label{eq.yfit}
\end{equation}
 The degree of the polynomial, $j_{max}$, is chosen according to
statistical criteria, ensuring that the fitted function provides a
good description of the values of the property in the whole range
$x \in[0,1]$. Using the functions given in Eq. (\ref{eq.yfit}) the
excess properties for the volume or the enthalpy are computed as a
function of $x$  as:
\begin{equation}
\bar{Y}^{E}(x,T,p) = \bar{Y}_f(x,T,p) - x \bar{Y}_f(1,T,p) - (1-x)
\bar{Y}_f(0,T,p).
\end{equation}

\subsubsection{The $\lambda_W=\lambda_S=0.25$ case}

First, we analyze the case in which the $B$-$C$  solvent-solute and
solute-solute interactions are both attractive and they have the same
value namely $\lambda_W=\lambda_S=0.25$. This  represents a system in
which in addition the solvent interacts with the solute in two
different ways, $B$-$A$ and $B$-$C$, both attractive. In principle
this would be the case of the water - alcohol mixture  where water
forms hydrogen bonds with the alcohol and shows and effective
attraction with the alkyl group.

\begin{figure}[!htb]
\vspace{0.7cm}
\includegraphics[clip,width=8cm]{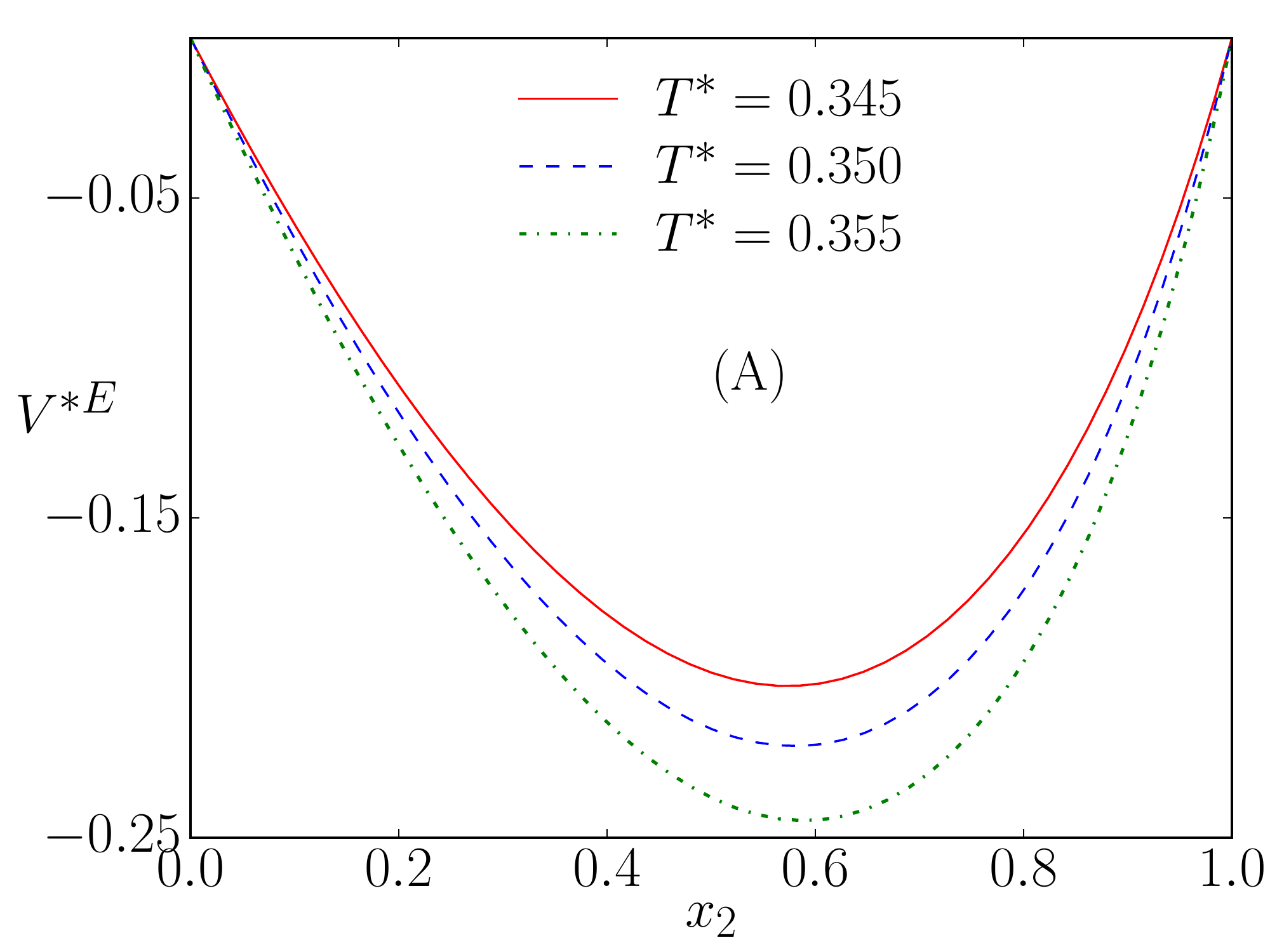}
\includegraphics[clip,width=8cm]{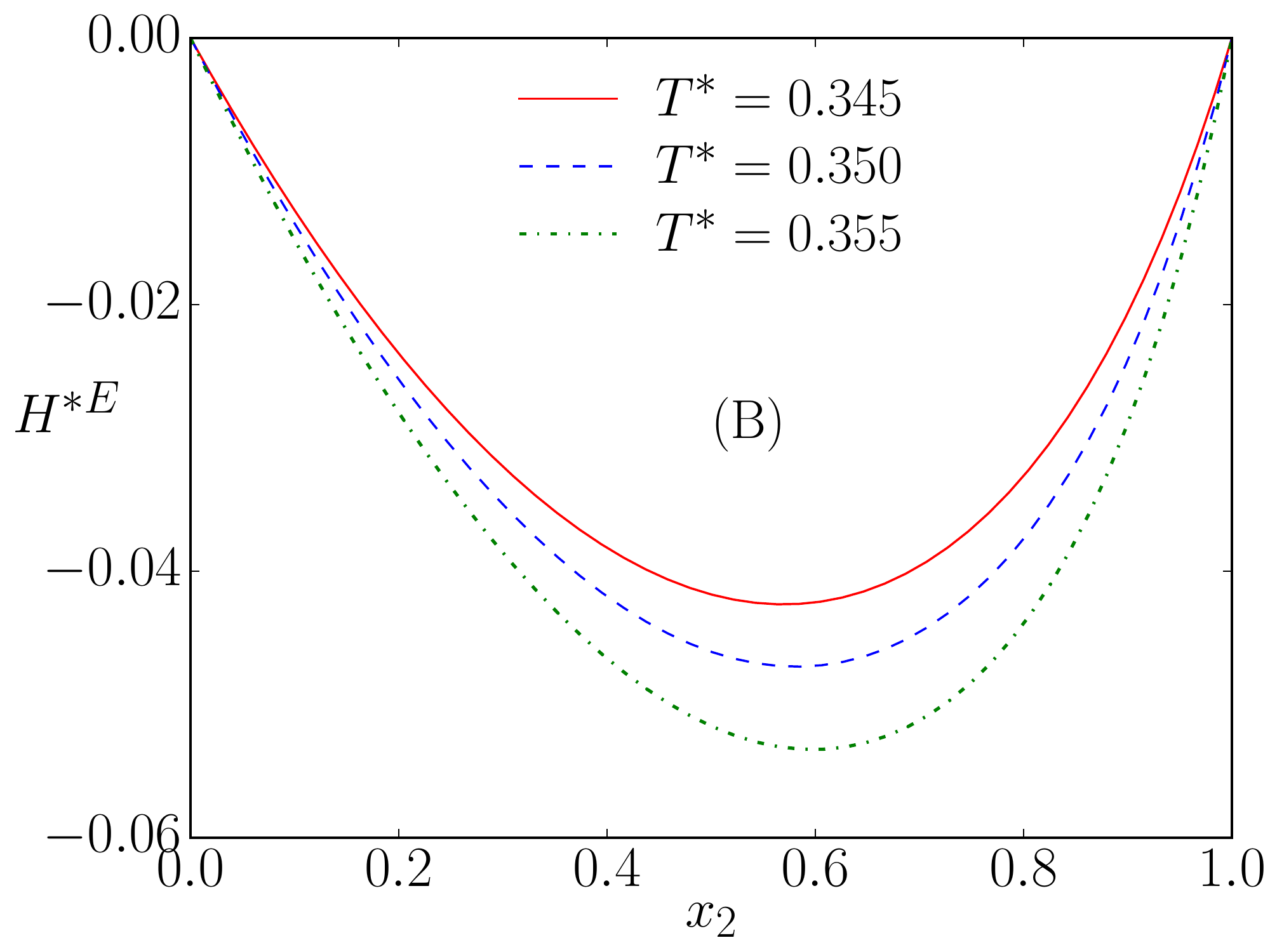}
\caption{ (A) excess volume and  (B) excess enthalpy
per particle  as  a function of solute concentration for
  $\lambda_{S}=0.25$,$\lambda_{W}=0.25$, $p^*=0.10$ and several
  temperatures.}
\label{fig.excess}
\end{figure}

The figure~\ref{fig.excess} illustrates the excess volume for the
pressure and temperature  $p^*=0.10$ and $T^*\simeq 0.3$. As the
fraction of the solute increases, the excess volume decreases until
it reaches a minimum. The presence of this minimum in a water-solute
 system is observed in the water-methanol
~\cite{Abello1973a,Benson1963a,Benson1980,Panov1976a,Pattel1985}, in
the water-ethanol~\cite{Ott1993}, in the
water-alkanolamines~\cite{Hepler1994,Stec2014} and in the
water-hydrophilic ionic liquids~\cite{Miaja2009} solutions.

The figure~\ref{fig.excess} also shows the excess enthalpy for our
model as a function of the fraction of the solute. The minimum
observed in $H^E$ is also present in the in
water-methanol~\cite{Tomasziewicz1986}, water-ethanol~\cite{Ott1986},
water-alkanolamines-\cite{Mundhwa2007} and in water with hydrophilic
ionic liquids~\cite{Miaja2009} solutions.

\begin{figure}[!htb]
  \includegraphics[clip,width=5cm]{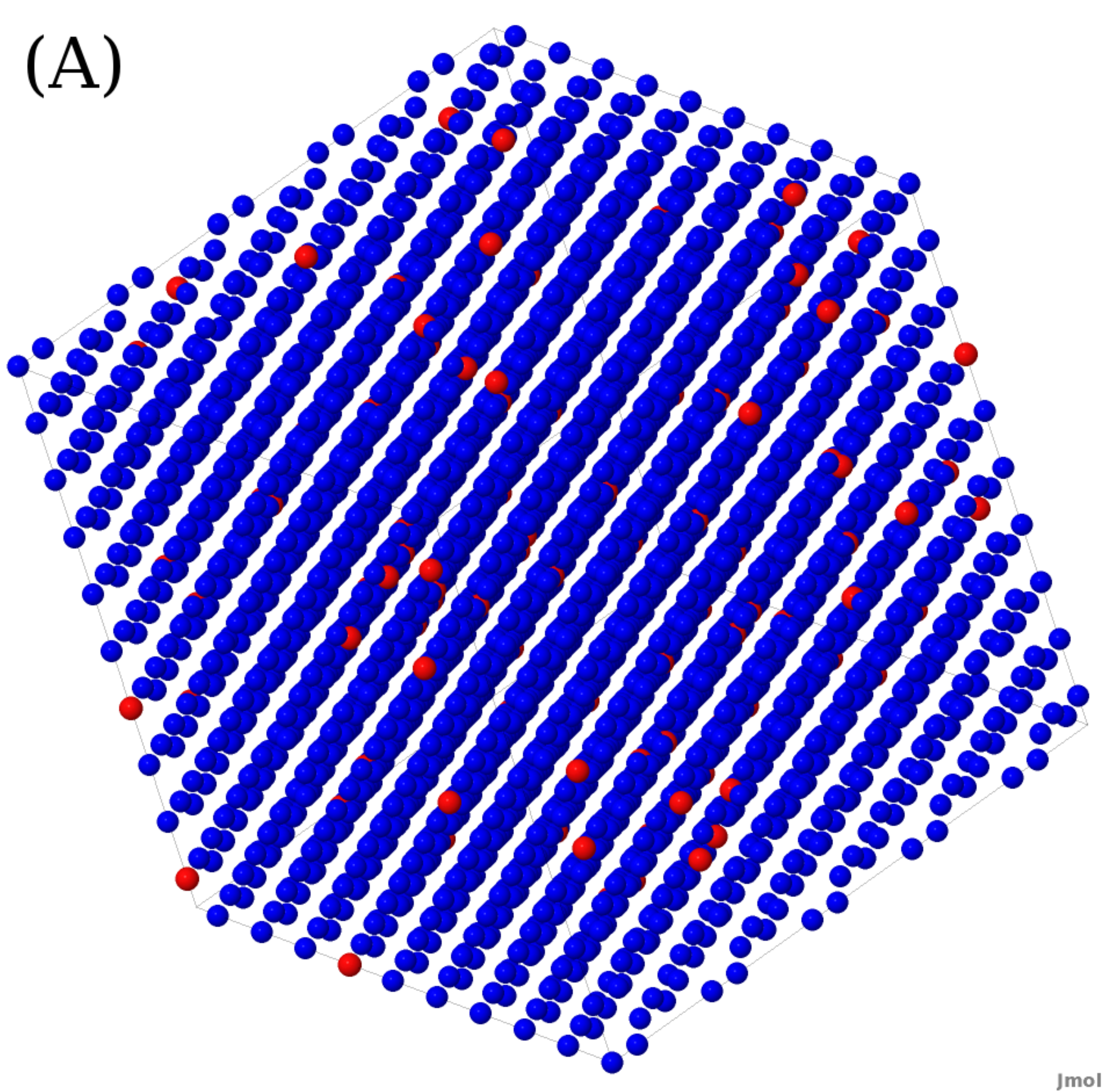}
\includegraphics[clip,width=5cm]{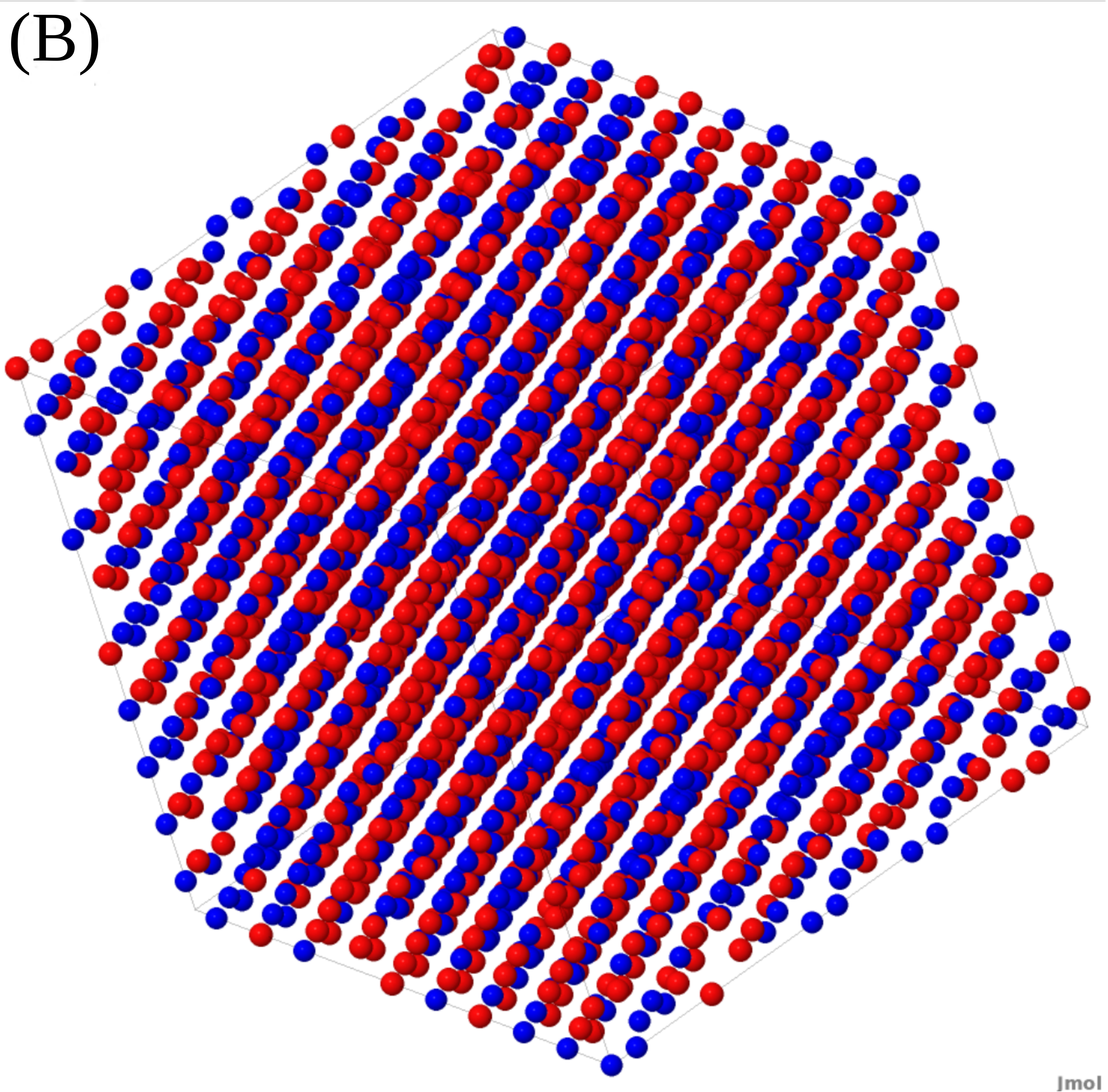}
\includegraphics[clip,width=5cm]{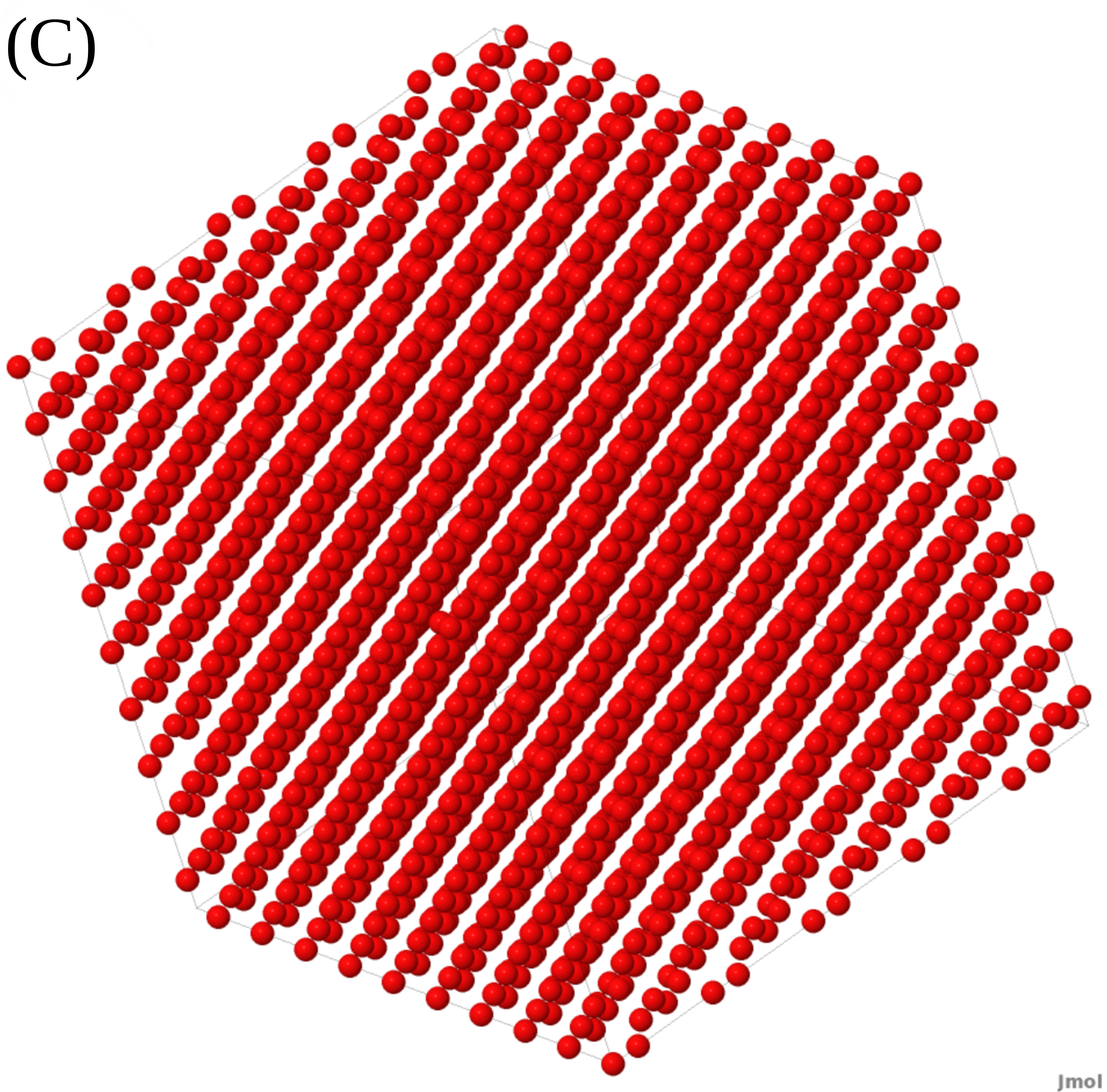}
\caption{The snapshot of system for $\lambda_W=0.25$ and
  $T^*=0.350$. The blue and red spheres represent the water and
  solute respectively. At (A)-$x_{2}=0.024105$,
   (B)-$x_{2}=0.704341$ and (C)-$x_{2}=1.$}
\label{fig:snap}
\end{figure}

The  figure~\ref{fig:snap} illustrates snapshots of the system as the
concentration of the solute is increased. Since the system is in the
LDL phase of  the solvent, there is one sublattice empty while the
other is filled. In the case in which the solute is a hard sphere, as
the solute is added to the system it enters in the empty sublattice
not competing with the solvent occupation~\cite{Szortyka2012}. Here
this is not the case. The solute enters in the same sublattice
occupied by the solvent.

The behavior of excess volume for this case is similar to that found
in quasi-ideal~\cite{Fuji2012} mixtures of equal size Lennard-Jones
particles and different potential well depth. In this quasi-ideal
mixture the Lennard-Jones energy parameters are given by,
$\epsilon_{22}/\epsilon_{11}=1.50$,
$\epsilon_{12}=\prt{\epsilon_{11}+\epsilon_{22}}/2$, therefore
$\epsilon_{12}/\epsilon_{11}=1.25$. At temperature below the critical
points and low pressure this mixture have negative excess volume and
its behavior as a function of the temperature is qualitatively equal
to that of our system. In relation to the excess enthalpy, the same
agreement occurs between the quasi-ideal~\cite{Fuji2012} system and
our model with $\lambda_W=\lambda_S$. The excess enthalpy is negative
for entire the range of the mole fraction and the departure from the
ideal mixture behavior increases on heating.

In the same mixture if we consider the Lorentz-Berthelot rule
$\epsilon_{12}/\epsilon_{11}=\sqrt{\epsilon_{22}/\epsilon_{11}}
\simeq 1.225$ (less favorable cross interaction) the excess volume
remains negative, but the excess enthalpy becomes positive at
temperatures slightly below the critical temperature of component one.

\begin{figure}[!htb]
 \includegraphics[clip,width=8cm]{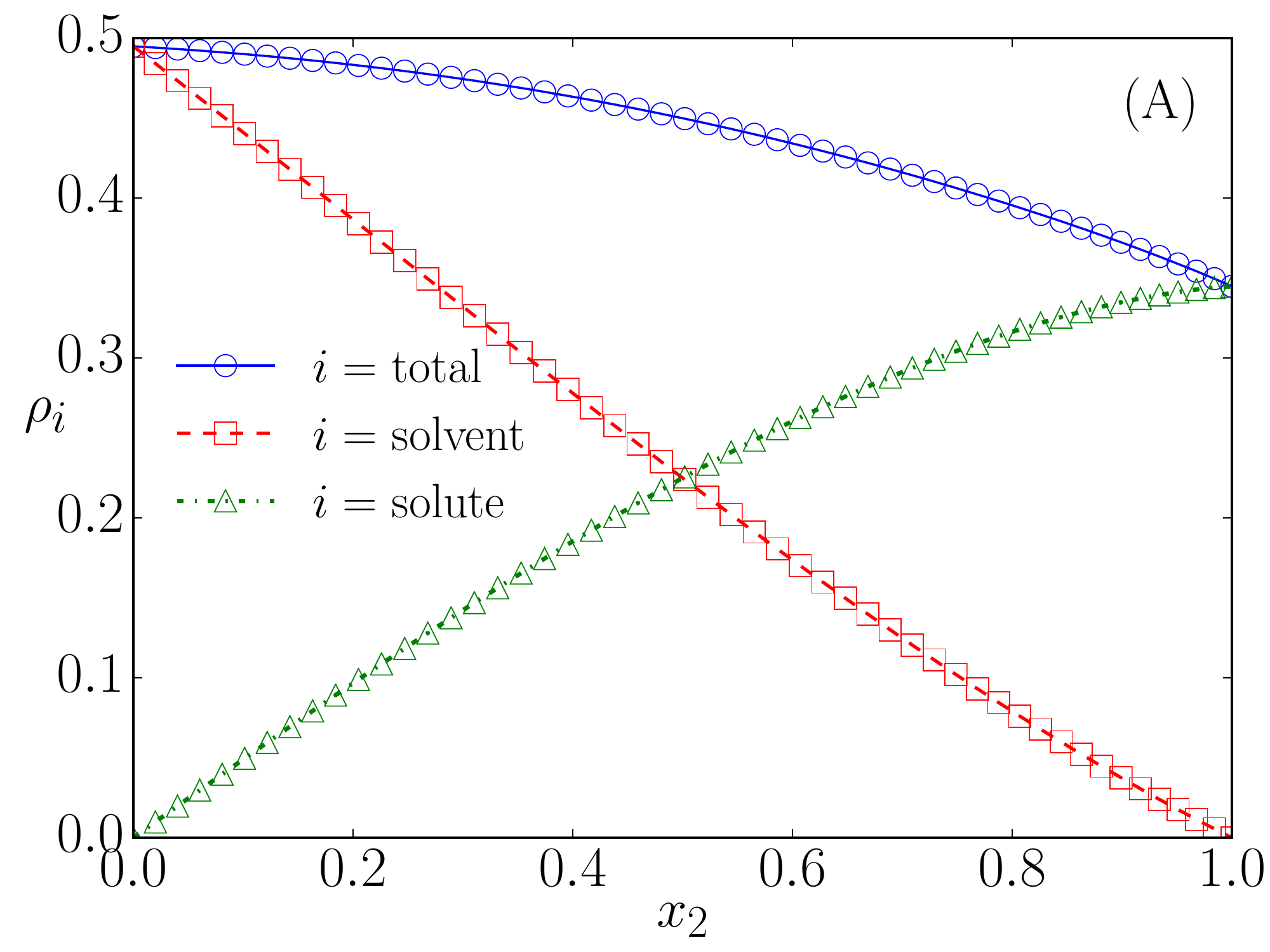}
\includegraphics[clip,width=8cm]{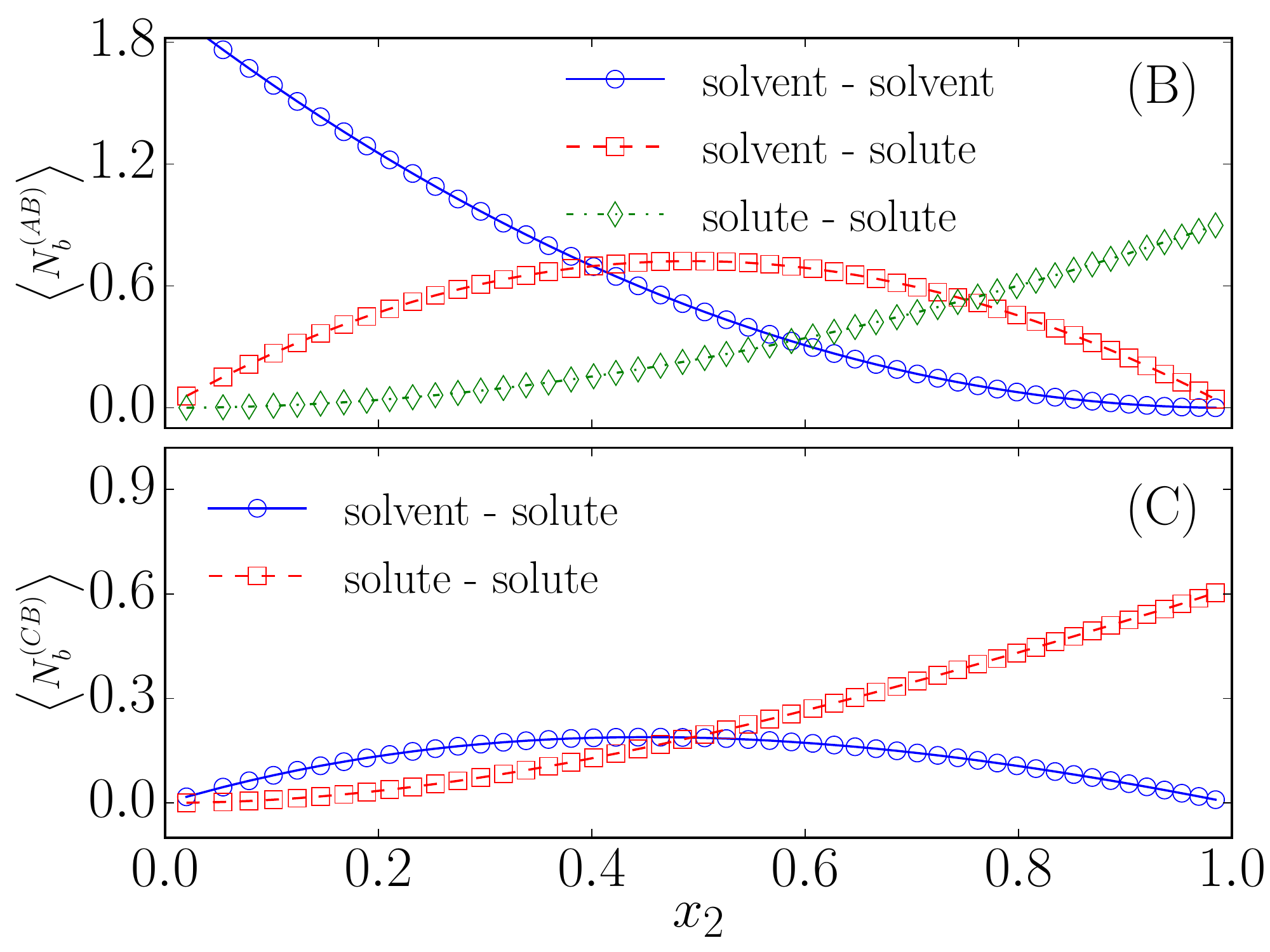}
\caption{(A) Total density and partial densities and average number of
  (B) $A$-$B$  bonds between solvent-solvent, solute-solute and
  solvent-solute and (C) $C$-$B$ bonds between solvent-solvent and
  solvent-solute for $\lambda_{M}=\lambda_{W}=0.25$, $p^*=0.10$ and
  $T=0.355$.at  $T^*=0.355$ and $p^*=0.10$.}
\label{fig:dens_mix}
\end{figure}

\subsubsection{The $\lambda_W=0$ and  $ \lambda_S=0.25$ case}

Next, we analyze the case in which the $B$-$C$   solute-solute  patch
is attractive but the $B$-$C$   solvent-solute patch  has no
interaction. In this case $\lambda_W=0$ while the solute-solute $B$-$C$
is attractive namely $ \lambda_S=0.25$. This  represents a system in
which in addition the solvent interacts with the solute only through
the $B$-$A$ patch. In principle, this would be the case of the
water-alcohol mixture in which the alkyl group is larger than the
preceding case and therefore the molecule is less hydrophilic.

\begin{figure}[!htb]
\vspace{0.5cm}
\includegraphics[clip,width=8cm]{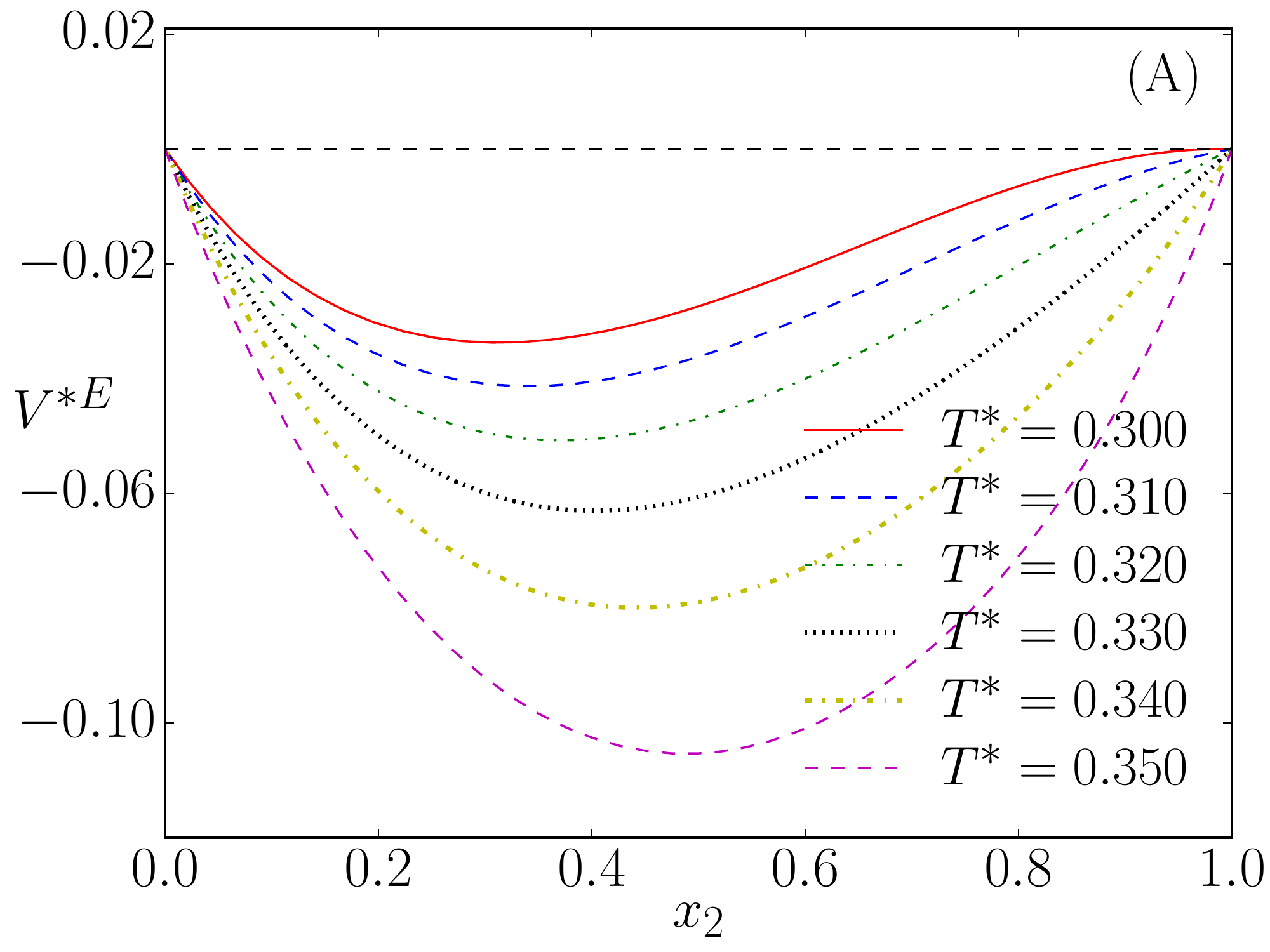}
\includegraphics[clip,width=8cm]{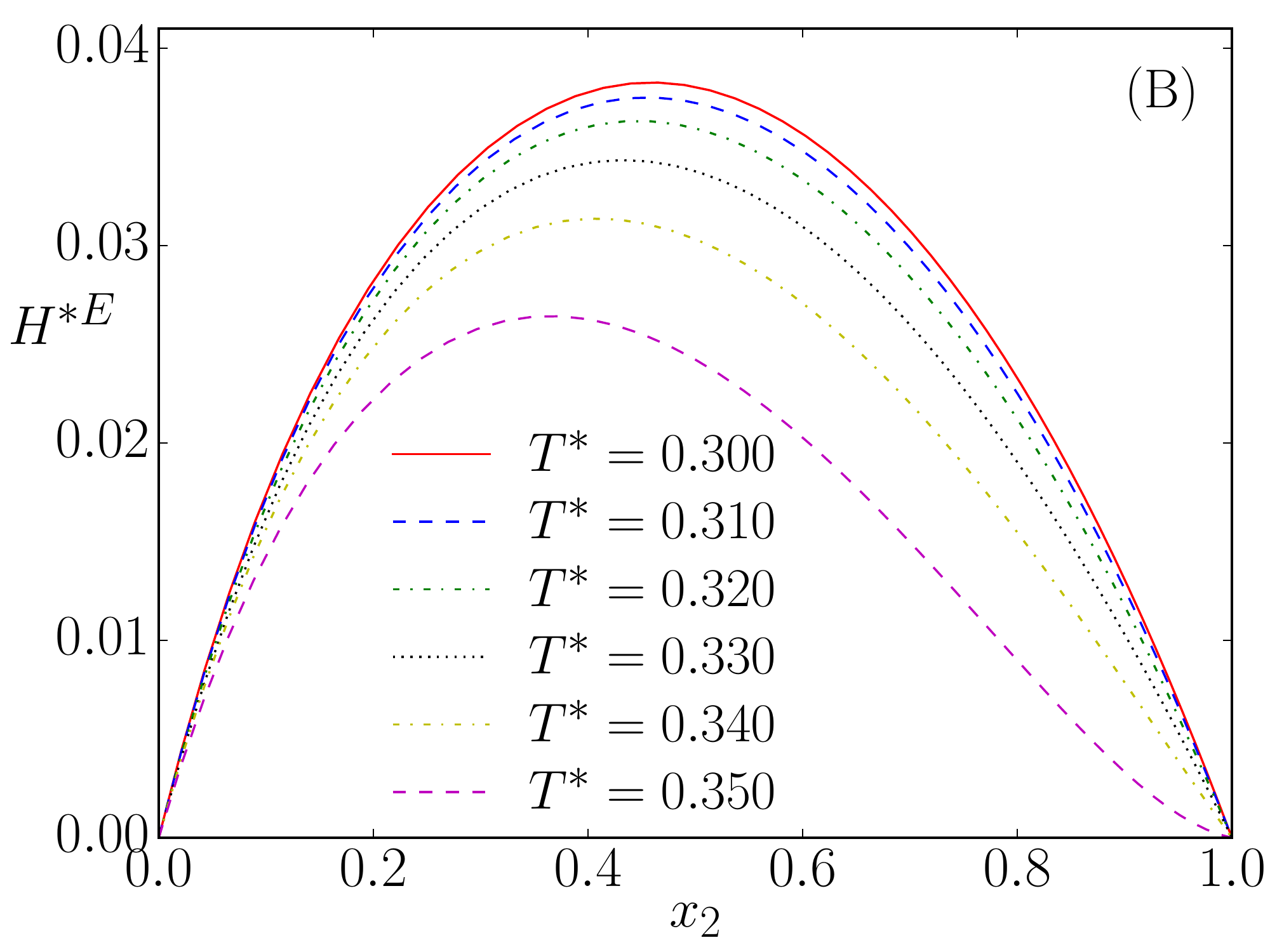}
\caption{(A) Excess volume and (B) Excess enthalpy $versus$
fraction of solute for various temperatures,  $\lambda_{W}=0$,
$\lambda_{M}=0.25$ and $p^*=0.10$.}
\label{fig.excess2}
\end{figure}

The figure~\ref{fig.excess2} illustrates the excess volume and the
excess enthalpy as a function of the fraction of the solute for
various temperatures. As the fraction of the solute increases, the
excess volume decreases until it reaches a minimum while the excess
enthalpy has a maximum. This behavior is consistent with the excess
volume and enthalpy of mixtures of water and large alcohol molecules
such as propanol, butanol and
pentanol~\cite{Ben-Naim2008,Marongiu1984}.

The variation of excess properties of this case with respect to the
previous one can be explained in terms of the fact that now cross
interactions are less attractive, which produces an increase in the
excess volume and enthalpy~\cite{Fuji2012}.

\subsubsection{The $\lambda_W=-0.25$ and  $ \lambda_S=0.25$ case}

Finally, we analyze the case in which the $B$-$C$ solute-solute patch
is attractive but the $B$-$C$ solvent-solute patch is repulsive. In
this case $\lambda_W=-0.25$ is repulsive while the solute-solute
$B$-$C$ is attractive namely $ \lambda_M=0.25$. This represents a
system in which in addition the solvent interacts with the solute
through the $B$-$A$ with attraction probably forming hydrogen bonds
while show repulsion through the $B$-$C$ patch. In principle this
would be the case of the water mixing with molecules that exhibit a
hydrophilic region, and eventually, can form a hydrogen bond, but the
overall water-solute interaction is repulsive.

\begin{figure}[!htb]
\vspace{0.5cm}
\includegraphics[clip,width=8cm]{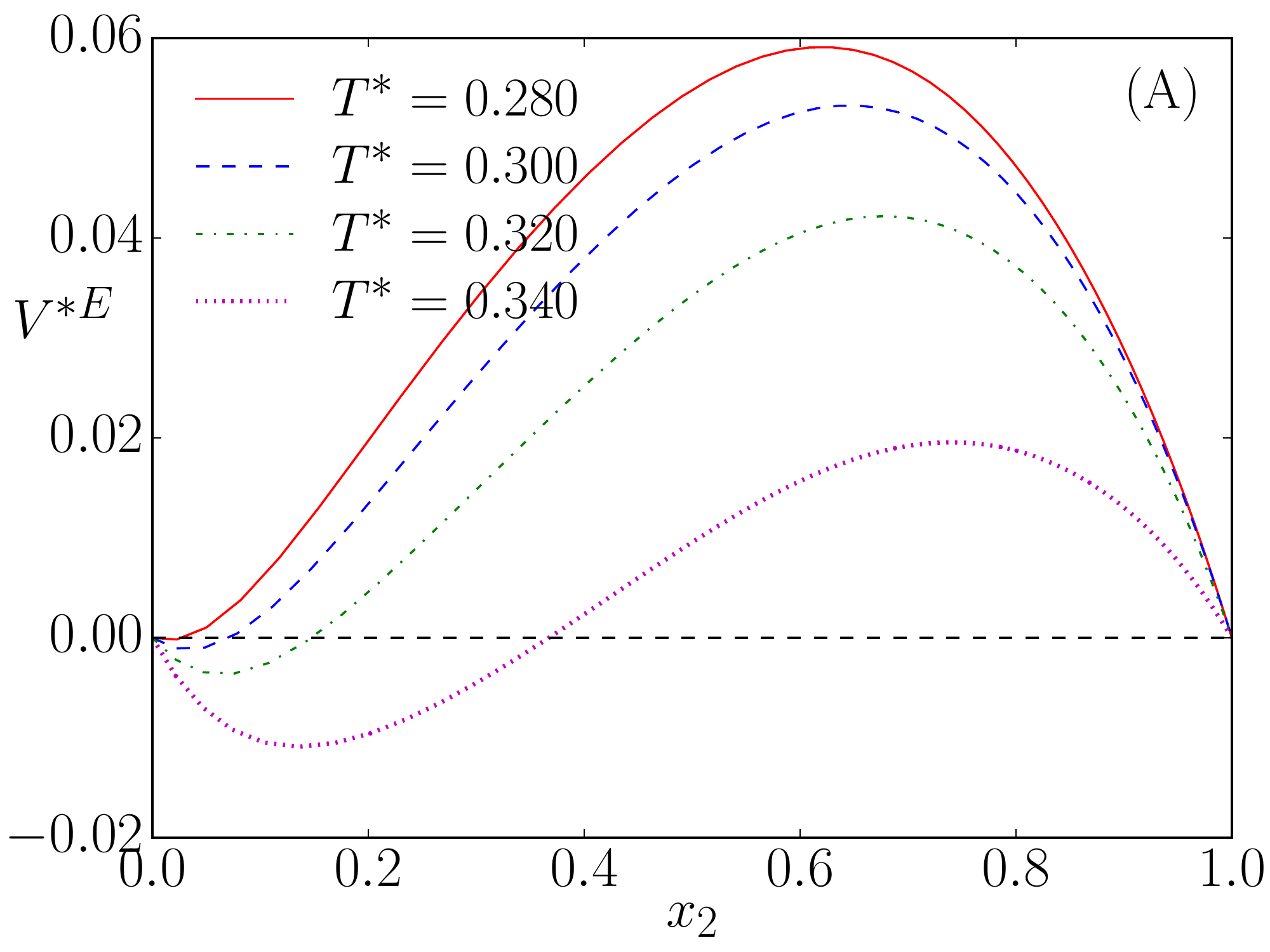}
\includegraphics[clip,width=8cm]{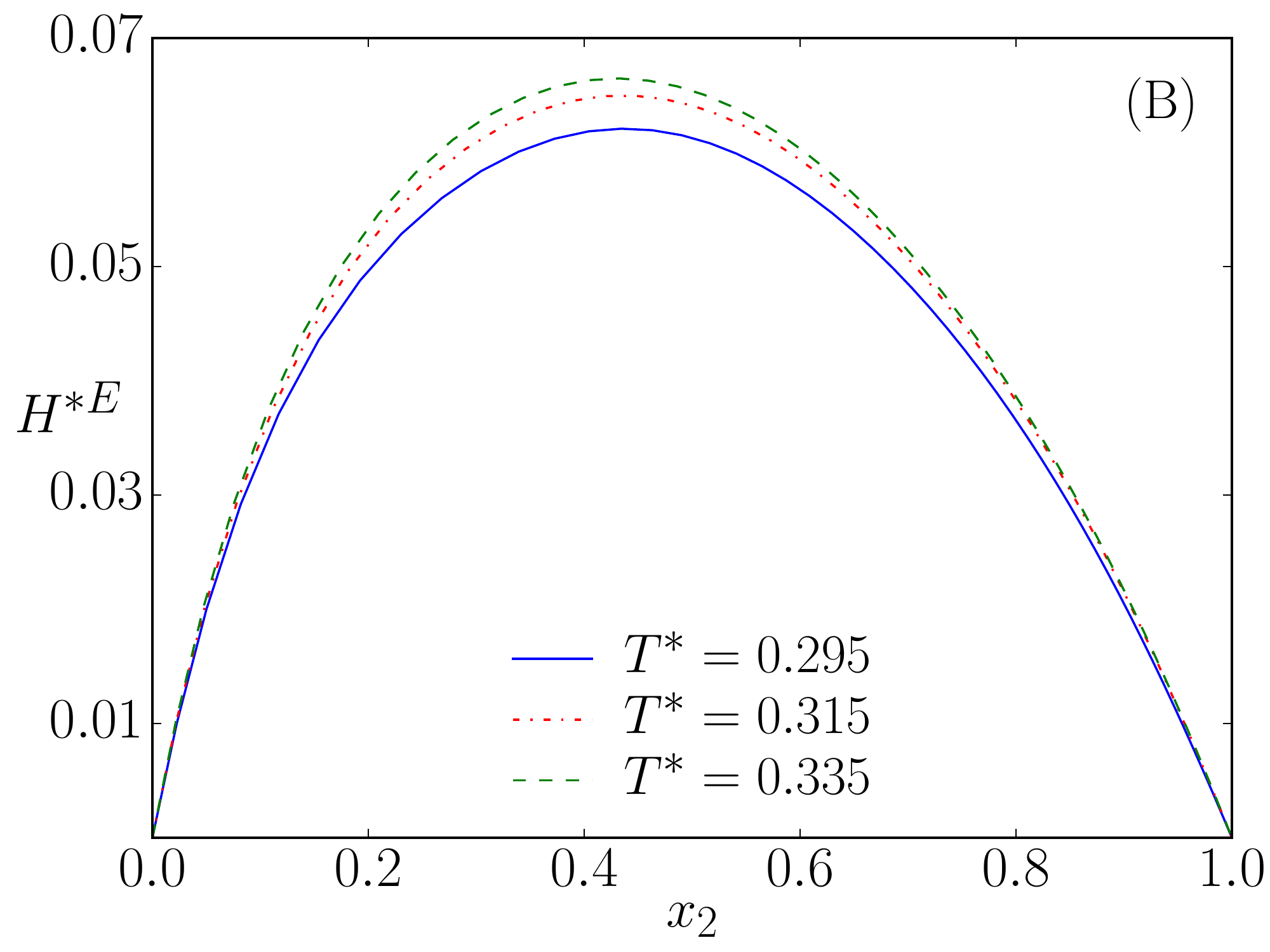}

\caption{(A) Excess volume and  (B) excess enthalpy $versus$ fraction of methanol for various temperatures.
$\lambda_{W}=-0.25$, $\lambda_{M}=0.25$ and $p^*=0.10$.
}
\label{fig.excess3}
\end{figure}

The figure~\ref{fig.excess3} illustrates the excess volume and the
excess enthalpy as a function of the fraction of the solute for
various temperatures. As the fraction of the solute increases, both
the excess volume and the excess enthalpy increase until
they reach a maximum. This behavior is found in hydrophobic ionic
liquids~\cite{Miaja2009,Vatascin2015}.

In this case, the cross interactions are more unfavorable and the
excess enthalpy and volume are both positives.

\begin{figure}[!htb]
  \includegraphics[clip,width=8cm]{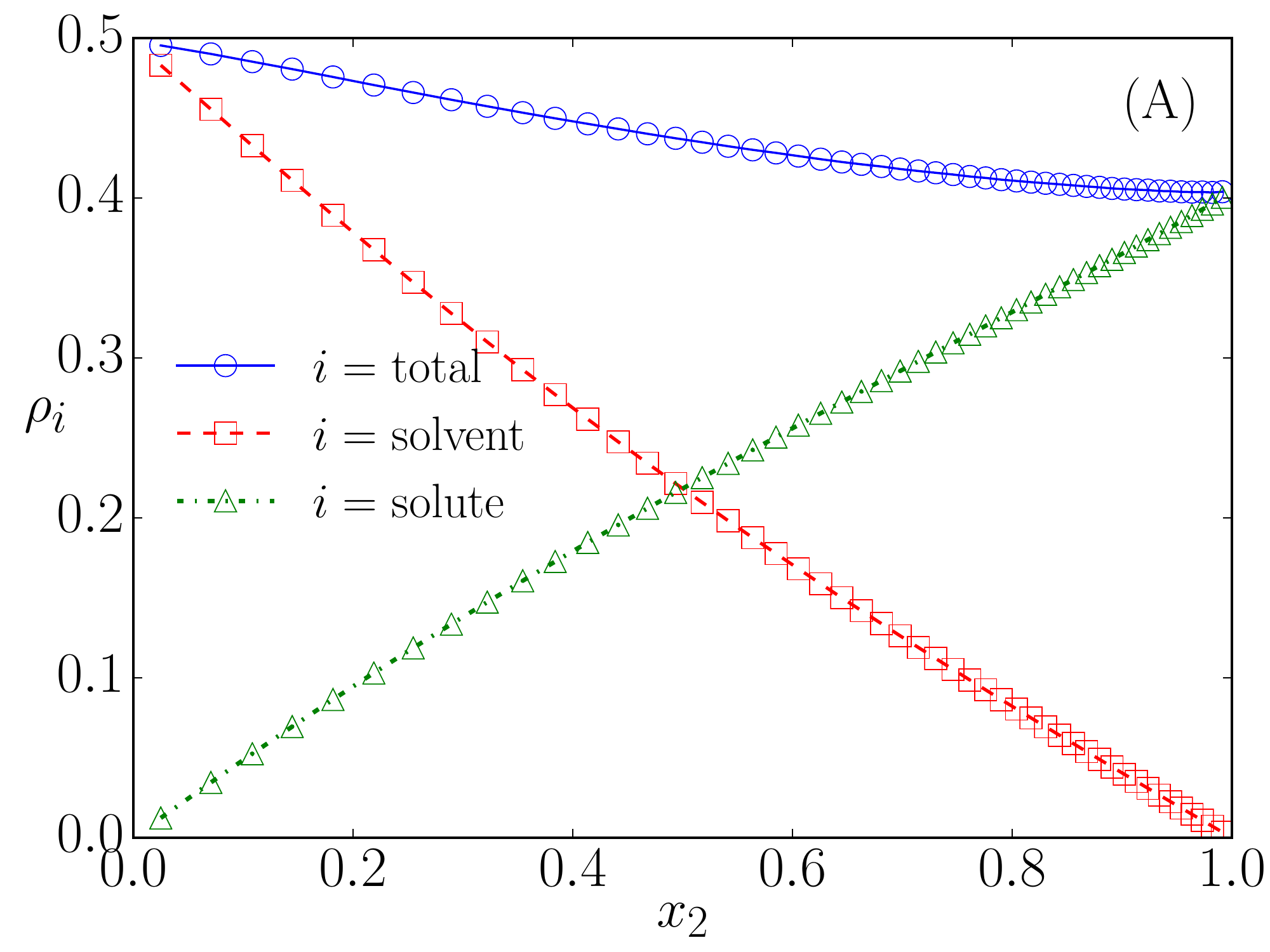}
  \includegraphics[clip,width=8cm]{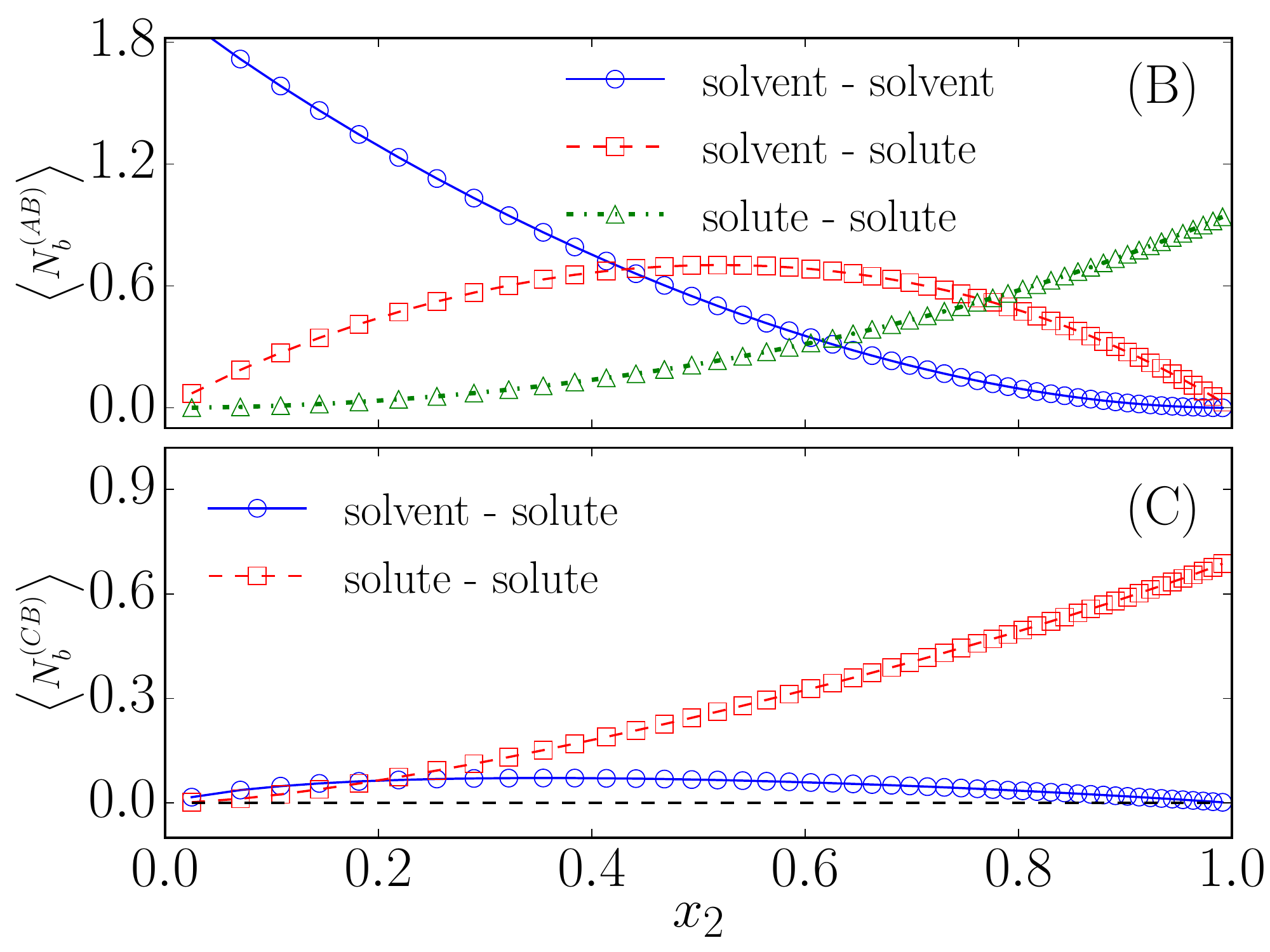}
\caption{(A) Total density and partial densities and average number of
(B) $A$-$B$  bonds between solvent-solvent, solute-solute and
solvent-solute and (C) $C$-$B$ bonds between solvent-solvent and
solvent-solute for $\lambda_{W}=-0.25$, $T^*=0.315$ and $p^*=0.10$.}
\label{fig:dens_mix2}
\end{figure}

\section{Conclusions}\label{sec:conclusions}

In this paper a combination of two Associating Lattice Models is
employed to represent a mixture of solute and solvent. In the
principle the solvent is modeled as a water-like system that exhibits
the density and diffusion anomalous behavior present in water. The
solute is modeled by molecules  that form two types of bonds with
water: one attractive hydrogen bond-like, the $A$-$B$ interaction,
plus an additional, tunable, $B-C$ interaction.

The pure components, by construction,  present very similar phase
diagrams. Both present, first (G-LDL/LDL-HDL) and  second order
($\lambda-$line/$\tau-$line) transitions, as well as some
multicritical points. The phases in coexistence are also the same. The
 substantial difference happens in the coverage area of phases and the
 position of multicritical points, and critical lines. The continuous
transitions belong to the 3D Ising model universality class.

For the mixture, by tuning the  $B$-$C$ patch from attractive to
repulsive we were able to qualitatively reproduce the behavior of
the excess volume and enthalpy of different types of mixtures.

The mapping of the relative hydrophobicity of the solutes through the
$\lambda_W$ parameter of the model, allows us to explain the trends
of the excess properties of the mixtures as function of the
intermolecular effective interaction.

Our result even though based on a very simple model reproduce a
mechanism that seems to be present in a large variety aqueous
solutions.

\appendix
\section{Computation of the residual entropy}\label{ap:entropy}

We consider a model defined over the diamond lattice with full
occupancy. Each particle (site) carries four arms which point to the
four NN sites. In order to compute the residual entropy of the lattice
 model for water we consider two arms of type $A$ (or +1) and two arms
 of type $B$ (or -1).
A given particle on the lattice can present $q_{W}  = 4!/(2! 2!) = 6 $
possible configurations. For the lattice model of solute, we consider
two arms of type $A$, two of type $B$ and one of type $C$, which leads
 to $q_M = 4!/2! = 12$ possible orientations of the particle.
We set interactions between NN particles, so that the interaction
energy is equal to zero for configurations compatible with the ground
state of the full model, and greater than zero another case. The pair
 interactions between NN particles are therefore $u=0$ when are due to
 the interactions between pair of arms $AB$ or $CB$, and
$u=\epsilon > 0$ for the other cases ($AA$, $BB$, $AC$, and $CC$).

The partition function of the system can be written as:
\begin{equation}
Q = \sum_{i=1}^{q^N} \exp \left[ -\frac
{U \left(\{ {\bf S} \}_i \right) }{ k_B T })\right]
\end{equation}
where $N$ being the number of particles (sites) of the system, $U$ the
 interaction energy and $q$ the number of possible orientations of
each particle. $\{ {\bf S} \}_i$ represent the different
configurations of the system. The Helmholtz energy, $A$, is related
with the partition function through:
\begin{equation}
\frac{A }{k_B T}= - \ln Q
\end{equation}
In the limit  of infinite temperature, all the possible configurations
 have the same probability and we get:
$A/(k_B T)  = -N \ln q$.
We are interested in the limit at low temperature. Given the fact that
 we know the partition function at high temperature, we can make use
of thermodynamic integration\cite{AllenTildesleyBook,FrenkelSmitBook}
to get:
\begin{equation}
 A(T) /(k_B T) = - N \ln q +  \int_{0}^{1/(k_B T) } U (T') d \left( \frac{1}{k_B  T'} \right) ;
\end{equation}
Taking into account the thermodynamic relation $A = U - T S $, $S$
being the entropy, and defining reduced quantities $U^*=U/\epsilon$,
and $T^*=k_B T/\epsilon$, we can get:
%
%
%
\begin{equation}
\frac{S(T)}{k_B}  = N\ln q_1 + \frac{U^*(T)}{T^*}- \int_{0}^{1/T^*}
U^* (T') {\textrm d} \left(1/T'^* \right).
\end{equation}
In the limit of low temperature the potential energy of the model
vanishes, and it is also fulfilled $\lim_{T\rightarrow 0} (U/T) = 0 $,
therefore we get:
\begin{equation}
\frac{S(T)}{k_B} \simeq N \ln q - \int_{0}^{1/T^*} U^*(T')
{\textrm d} \left(1/T'^*\right)  ; \quad ( T^* \rightarrow 0 )
\end{equation}
The residual entropy per particle $s_0(N)$ (as a function of the
system size) can be computed as:
\begin{equation}
\frac{s_0(N)}{k_B}  = \ln q - \lim_{T^* \rightarrow 0}
\int_{0}^{1/T^*} \frac{U^* (N,T')}{N} \textrm{d} \left( 1/T'^* \right)
\end{equation}
The determination of $s_0(N)$ has been carried out using Monte Carlo
simulation, in combination with thermodynamic integration, and
parallel tempering techniques. Different system sizes were considered
in order to carry out a finite-size scaling analysis to determine
$s_0$ in the thermodynamic limit. Parallel tempering facilitates the
equilibration of the systems at low temperature, where the systems
reach the ground state (except for some elementary excitations).
For the lattice model for water, we have considered different system
sizes: $N=8 \ell^3$, with $\ell = 2, 3, 4, \cdots,  14$. In each case
we considered 257 values of $(1/T^*)$;
$1/T_i^*= i\times \Delta (1/T^*); \quad i=0, 1,\cdots ,256$;  with
$\Delta (1/T^*) = 0.050$. The averaged reduced potential energy per
particle $u^* = U/(N\epsilon)$ is $1$ for $T\rightarrow \infty$, and
it almost vanishes for the lowest values of $T$ considered in the
integration $u^* \ll 10^{-6}$ (for the largest system sizes). It decays
 rapidly as $T \rightarrow 0$, making possible a reliable cut-off of
the integration for a given level of accuracy in the results.
In TABLE~\ref{table-s0} we present the estimates for $s_0(N)$.
\begin{table}
\caption{System-size dependent estimates for the residual
entropy presented as $s_0^* (N)/k_B$}
\begin{tabular}{||c|ccccc||}
\hline \hline
$\ell $ & 2 & 3  & 4 & 5  & 6 \\
\hline
$s_0(8\ell^3)/k_B$ &  0.435774(13) & 0.418939(18) & 0.414306(18) &
 0.412543(16)  &   0.411693(18) \\
\hline  \hline
$\ell $ & 7 & 8  & 9 & 10 & 11\\
\hline
$s_0(8\ell^3)/k_B$ & 0.411271(19) & 0.410988(22) &  0.410823(21)
&  0.410737(24) & 0.410645(21)
\\
\hline  \hline
$\ell $ &12 &13  &14 & & \\
\hline
$s_0(8\ell^3)/k_B$ & 0.410619(14) &   0.410574(13) & 0.410549(8) & &
\\
\hline
\hline
\end{tabular}
\label{table-s0}
\end{table}
In order to estimate the value of $s_0$ in the thermodynamic limit we
have considered the scaling relations used by
Berg {\it et al.}~\cite{Berg2007},
\begin{equation}
s_0(N)/k_B = s_0/k_B + a_1 N^{-\theta};
\label{Eq-s0m}
\end{equation}
The fitting of the simulation results given in TABLE \ref{table-s0}
to Eq. (\ref{Eq-s0m}), with $(s_0/k_B)$, $a_1$, and $\theta$ being
adjustable parameters leads to:
\begin{equation}
s_0^{\textrm{(W)}}/k_B =
0.410\; 41 \pm 0.000\;  02; \; \; \;
\theta = 0.899 \pm 0.005,
\end{equation}
where the label $(W)$ refers to water. Considering the quantities
$\Omega(N_L)=\exp[s_0(N_L)/k_B]$, and fitting the results to
\begin{equation}
\Omega(N) = \Omega + a_{\Omega} N^{-\theta},
\label{eq-wm}
\end{equation}
we get
\begin{equation}
\Omega= 1.507\; 44 \pm 0.000\;  04; \; \; \;
\theta = 0.905 \pm 0.005 .
\end{equation}
The values of the exponent $\theta$ agree within statistical
uncertainty with the results of Berg et al.~\cite{Berg2007}. For the
residual entropy of the ordinary ice. Interestingly, our estimate of
$\Omega $ for our model defined over a system with cubic symmetry and
the estimate of for the ordinary ice of Berg et al.~\cite{Berg2007}:
$\Omega^{\rm Ice} = 1.507 \; 38 \pm 0.000 \; 16$; $\theta = 0.923(23) $,
seem to coincide (at least within error bars) in spite of the different
structures of the underlying lattices.

In principle, we could apply the same simulation techniques used for
the water in the determination of the residual entropy of the lattice
gas model of the solute.
However, the value of $s_0$ for methanol can be deduced directly from
the water results. Given a ground state, the configuration of the
water for a system with $N$ molecules (occupied positions) one can
build up $2^N$ directly related ground states for the methanol model,
since the two (undistinguishable) $A$ patches of each particle in the
 water model correspond to two distinguishable ($A$ and $C$) patches
in the methanol model.
Therefore, we get:
\begin{equation}
s_0^{\textrm{(S)}}  =
s_0^{ \textrm{(W)}} + k_B \ln 2.
\end{equation}

\section{Computation of  isobars for pure components}
\label{sec.isop}

The excess properties of binary mixtures are usually measured
experimentally at fixed conditions of temperature and
pressure~\cite{Marongiu1984,Pattel1985}.
For lattice gas models it is neither straightforward not practical the
 use of simulation in the NPT ensemble.
The usual alternative is to carry out simulations in the grand
canonical ensemble and compute the pressure by means of thermodynamic
integration.
Since we are interested in analyzing the excess properties at fixed
pressure, we have developed a procedure to build up the lines
$\mu(T|p)$ for pure components, i.e. we fix the pressure and compute
the chemical potential as a function of temperature at fixed pressure.
The objective is to apply this to the ordered phases: LDL and HDL. The
 pressure at (very) low temperature for these phases can be computed
from the ground state analysis.
In the GCE the change of the pressure for transformations at constant
$T$ and $V$, is given by $d p = \rho d \mu$.
The density of the condensed phases at very low temperature hardly
changes with $\mu$, therefore, we can integrate the pressure to get.
\begin{equation}
p = p_0 + \left( \mu - \mu_0 \right) \rho_0
\end{equation}
where the values of $p_0$, $\mu_0$, and $\rho_0$ can be taken as those
corresponding to the phase coexistence at low temperature
(Eqs. \ref{eq.gsw}-\ref{eq.coexm}).
Once we now how to compute the chemical potential for a given pressure
 $p$ at a (low) temperature $T_1$, we will develop the integration
scheme to move on the $(\mu,T)$ plane at the fixed pressure $p$.
Imposing $dp=0$ in the differential form for the thermodynamic
potential of the GCE we get:
\begin{equation}
d\mu = - \frac{ U + p V - N \mu}{ N T } d T = -
\frac{ \tilde{u} - p + \mu \rho}{\rho T} d T
\label{eq.dmudt}
\end{equation}
We typically considered systems with $N_L = 2 \times 16^3$.

\section{The properties of mixtures at fixed $T$ and $p$}
\label{sec.isopmix}
The excess properties of mixing are usually defined as the differences
 between the values of the property of the mixture at a given
composition,  $x$, and the value of the same property for an
{\em ideal} mixture of the components at the same conditions of $x$,
$T$, and $p$. It is, therefore, desirable to develop simulation
strategies to sample in an efficient way different compositions of a
given mixture for fixed conditions of temperature and pressure. In
order to achieve this aim for our lattice model we have borrowed ideas
to form the Gibbs-Duhem integration procedures, as we did for computing
isobars of pure components.

The differential form for the grand canonical potential of a binary
mixture can be written as:
\begin{equation}
  - d \left( \frac{pV}{T}\right) = U d \left( \frac{1}{T} \right)-
\frac{p}{T}d V -  \sum_{i=1}^2 N_i d \left( \frac{\mu_i}{T} \right);
\end{equation}
where $N_i$ is the number of molecules of component $i$, and $\mu_i$
is the chemical potential of component $i$. If we fix $T$, $p$, and
$V$, the chemical potential of the two components can not vary
independently when modifying the composition. It should be fulfilled:
\begin{equation}
N_1 d \mu_1 + N_2 d \mu_2 = 0 \; .
\label{gdi}
\end{equation}
Using activities $z_i \equiv \exp[ \mu_i/(k_B T)]$ to carry out the
integration of Eq. (\ref{gdi}) we get:
\begin{equation}
\frac{N_1}{z_1} d z_1 +
\frac{N_2}{z_2} d z_2 = 0\; .
\label{gdi2}
\end{equation}

Let us assume that for some values of $T$, and $p$, we know the values
 of the activities of the pure components $z_1^{(0)}$, and
$z_2^{(0)}$. We can integrate numerically (using simulation results)
the differential equation:
\begin{equation}
d z_2 =- \frac{N_1 z_2}{N_2 z_1} d z_1\; .
\label{eq.gdi3}
\end{equation}
For instance, using as starting point  $(z_1=z_{1}^{(0)},z_2=0)$ and
considering $z_1$ as the independent variable and integrating Eq.
(\ref{eq.gdi3}) up to $z_1=0$, we should reach $z_2(z_1=0)=z_2^{(0)}$.
 This condition provides a powerful consistency check of the
thermodynamic integration schemes at constant pressure. The numerical
integration of (\ref{eq.gdi3}) can be carried out using the same
numerical procedures as in Sec.~\ref{sec.isop}. There is still,
 a minor technical problem, that appears in the limits
$z_i \rightarrow 0$; where $N_i\rightarrow 0$, and therefore the ratio
 $(N_i/z_i)$ can not be directly computed from the simulation.
This problem can be solved by applying the Widom-insertion test
technique\cite{FrenkelSmitBook} to compute the activity of the
minority component (which actually has mole fraction $x=0$) as a
function of its density. The result can be written as:
\begin{equation}
\lim_{z_i \rightarrow 0^+} \frac{\rho_i^*}{z_i}  = q_i
\aver{\exp\cch{-\Delta u_{i}/(k_BT)}};
\label{eq.int-mix}
\end{equation}
where $\aver{\exp\cch{- \Delta u_i/(k_BT)}}$ represents the average of
the Boltzmann exponential over attempts of insertion of a test
particle of type $i$ with random position and random orientation on a
pure component system of the other component and $q_i$ is the number
of possible orientations for molecules of type $i$. Results were
obtained from simulations of systems with $N_L = 2 \times 16^3$.


\acknowledgments
A. P. Furlan and M. C. Barbosa acknowledge the Brazilian agency CAPES
(Coordena\c{c}\~ao de Aperfeicoamento de Pessoal de N\'ivel Superior)
for the financial support and Centro de F\'isica Computacional - CFCIF
 (IF-UFRGS) for computational support. Partial financial support from
the Dirección General de Investigación  Científica y Técnica (Spain)
under Grant No. FIS2013-47350-C5-4-R is acknowledged. In the course of
 writing this article, Noé G. Almarza unexpectedly passed away. The
authors would like to dedicate this work to his memory.


\end{document}